\documentclass[journal=jacsat,manuscript=article]{achemso}

\usepackage[version=3]{mhchem} 
\usepackage{xcolor}

\usepackage[pagewise]{lineno}

\usepackage{xr}
\usepackage{hyperref}
\usepackage{comment}

\makeatletter

\newcommand*{\addFileDependency}[1]{
\typeout{(#1)}
\@addtofilelist{#1}

\IfFileExists{#1}{}{\typeout{No file #1.}}
}\makeatother

\newcommand*{\myexternaldocument}[1]{%
\externaldocument{#1}%
\addFileDependency{#1.tex}%
\addFileDependency{#1.aux}%
}
\myexternaldocument{02_SI}

\SectionNumbersOn

\author{NP Vaisakh}
\affiliation{Soft Matter and Nanomaterials Laboratory, Department of Physics, Indian Institute of Technology Bombay, Mumbai-400 076, India}
\author{Suman Bhattacharjee}
\affiliation{Centre for Research in Nanotechnology \& Science (CRNTS), Indian Institute of Technology Bombay, Mumbai-400 076, India}
\author{Sunita Srivastava}
\email{sunita.srivastava@iitb.ac.in}
\phone{}
\fax{}
\affiliation
{Soft Matter and Nanomaterials Laboratory, Department of Physics, Indian Institute of Technology Bombay, Mumbai-400 076, India}

\title
  {Role of Fluid Forces and Depletion Interactions in Directing Assembly of Aqueous Gold Nanorods on Hydrophobic Surfaces\footnote{Preprint}}

\begin{document}

\begin{abstract}
The interaction between macroscopic fluid flow and nanoscale forces has resulted in the formation of long-range assemblies through evaporation-induced self-assembly. Anisotropic gold nanorods (AuNR) can form disordered, smectic, or vertically ordered long-range structures, but controlling their assembly remains a challenge and requires a deeper understanding of fundamental interaction mechanisms. In this work, we established a correlation between the \textit{in situ} drying profiles, measured using an optical tensiometer, and deposit pattern, imaged \textit{ex situ} using electron microscopy. Increasing particle concentration induced a transition from coffee-ring to uniform deposition at the microscale, while nanoscale structures shifted from isotropic/smectic to vertically aligned crystalline AuNRs. The interplay of capillary and Marangoni flow influences assembly at both macro and nanoscales, with the deposition process and nanoparticle ordering being highly sensitive to interparticle and nanoparticle-substrate interactions. By systematically studying key parameters, we aim to develop a comprehensive framework for the rational design and fabrication of nanomaterials with precisely controlled structure and properties.

\end{abstract}

\section{Introduction}
Evaporative self-assembly (ESA) emerges as an appealing choice among a spectrum of methods for generating patterned particle depositions, owing to its simplicity, scalability, and cost-effectiveness, particularly in creating colloidal films catering to diverse applications\cite{tang2017efficient,lohse2022fundamental,vialetto2024versatile,bhardwaj2020likelihood,apte2015vertically}. Solvent evaporation based assembly of nanomaterials over a flat surface has been reported to yield various surface structures in dried deposits, such as the ubiquitous coffee-ring \cite{deegan1997capillary,yunker2011suppression,mampallil2018review,anyfantakis2015manipulating,bhattacharjee2023ordered}, uniform depositions\cite{yunker2011suppression,anyfantakis2015manipulating,li2016rate}, concentric rings\cite{srivastava2020dual}, etc. Droplet evaporation is a complex, non-equilibrium process, and controlling the nanoparticle patterns formed during evaporation is highly desirable. However, achieving precise control remains challenging and has been less extensively explored. During evaporation of a colloidal droplet with non-volatile solute particles, it typically leaves behind a ring-like deposit known as the coffee-ring, as documented in Deegan's early works\cite{deegan1997capillary,deegan2000pattern}.
To counteract the uneven evaporation across the curved surface of a semi-spherical droplet, a radially outward capillary flow is induced within the droplet, dragging the solutes to the three-phase contact line (TPCL) and forming the coffee-ring.
Apart from capillary flow, solutal and thermal Marangoni flows, whose directions can be experimentally tuned, determine the resultant motion of the solute particles and hence the deposit pattern of the sessile droplet \cite{ristenpart2007influence,nguyen2002patterning,gelderblom2022evaporation}. \par
Although fluid flow affects the microscopic deposition patterns, considering nanoscale interactions between solute-solvent and solute-substrate is important and have been found to vary the morphology of surface deposits\cite{anyfantakis2015manipulating,li2016evaporative}. Much work has been done to suppress/reduce the coffee-ring formation by tuning the particle-particle and particle-interface interactions\cite{bhattacharjee2023ordered,anyfantakis2015modulation,khawas2023anisotropic}, decreasing the pinning forces and reducing the net capillary flow\cite{yunker2011suppression,anyfantakis2015manipulating,bridonneau2020self}.
The usage of low-wetting surfaces for solvent-based self-assembly of nanoparticles has resulted in distinct assemblies as compared to highly wettable surfaces\cite{bhattacharjee2023ordered,srivastava2020dual}. The hydrophobicity of the substrate affects the dynamics of the TPCL, with increased hydrophobicity promoting TPCL depinning.\cite{orejon2011stick} TPCL smoothly recedes when nanoparticles have a feeble affinity to the substrate, forming a compact dried deposition of dimensions much less than that of initial droplet diameter and affecting the surface morphology of the dried deposits \cite{li2016evaporative,mampallil2012control, marin2012building, seyfert2021evaporation}. 
\par
Anisotropic nanoparticles such as gold nanorods (AuNR) have garnered significant research interest\cite{zheng2021gold} thanks to their ability to organize into smectic\cite{ming2008ordered}, nematic\cite{umadevi2013large,liu2010self}, and vertical domains\cite{thai2012self}, marking a noteworthy stride towards fabricating SERS-based sensors\cite{peng2013vertically}, water treatment devices\cite{loeb2019nanoparticle}, storage devices\cite{zijlstra2009five}, etc. The orientation of nanorods can be controlled by varying parameters like surfactant concentration\cite{khawas2024directing}, temperature\cite{kruse2024temperature}, humidity\cite{zhang2014high}, wettability\cite{kim2014mediating} of the surface, etc. Solvent-evaporation-induced assembly of CTAB-coated AuNRs on a hydrophilic substrate formed a coffee-ring followed by the surfactant-filled depletion zone\cite{khawas2023anisotropic}. A region of small arrays of nanorods was reported outside the coffee-ring, which is related to the initial retraction of the TPCL due to the autophobic effect. On the other hand, hydrophobic/low-wetting surfaces has been reported to exhibit different patterns such as periodic ring structures\cite{li2016direct} and vertical superlattices\cite{li2016evaporative} of AuNRs. The formation of a coffee-ring pattern is attributed to the pinning effect and capillary flow, which organize AuNRs at the nanoscale into smectic long-range multilayers. However, the mechanism behind the formation of distinct AuNR assemblies on hydrophobic surfaces requires further investigation.

\par
In this study, we examine and analyze dried patterns created by the evaporation of colloidal droplets of CTAB-capped AuNRs at various concentrations on a hydrophobic substrate. The study places particular emphasis on examining how unfavorable particle-substrate interactions influence surface morphology, spanning from the microscale to the nanoscale. For this purpose, we employ solvent-evaporation-based self-assembly of CTAB-capped AuNRs on a low-wetting substrate, offering an alternative approach to suppress coffee-ring formation. A transition from a coffee-ring to a uniform deposition occurs when the AuNR concentration crosses a critical value. Drying profiles were recorded using an optical tensiometer, and the contrasting changes in evaporation profiles on hydrophobic substrates and their effects on the resulting dried patterns are discussed. The nanoscale AuNR assembly, measured \textit{ex situ} using scanning electron microscopy reveals the formation of distinct crystalline phases of AuNR, arranged in smectic and triangular lattices, along the coffee ring deposit. These phases were found to depend on the concentration of the colloidal suspension. Our study shows that lowering surface wettability results in the formation of large-scale crystalline domains with vertically aligned rods, especially at high AuNR concentrations. The roles of capillary and Marangoni flow, along with depletion interactions, are discussed in relation to the variable assembly from the micro- to the nanoscale. The experimental results presented here provide insights into the strong influence of nanoscale interactions on macroscale deposition patterns. These studies are important for several applications exploiting nanoparticle suspension for surface coating, e.g. for the creation of antimicrobial and self-cleaning surfaces\cite{wang2010preparation}. Furthermore, the creation of large areas of vertically oriented gold nanorods serves as hot spots of electromagnetic field and plays a crucial role in the development of SERS-based sensors. 

\section{Materials and methods}
\subsection{Reagents}
Hydrogen tetrachloroaurate trihydrate (\ce{HAuCl4 * 3H2O}, Sigma Aldrich, 99.99 \%), cetyltrimethylammonium bromide (CTAB), sodium borohydride (\ce{NaBH4}, Sigma Aldrich, 99 \%), ascorbic acid (Merck, 99 \%), silver nitrate (\ce{AgNO3}) were used for AuNR synthesis. Perfluorooctyl-trichloro silane (PFOTS) (Sigma Aldrich, 97 \%) was used for coating $Si$ surfaces. Nile red (Sigma Aldrich) and Dimethylsulphoxide (DMSO, Sigma Aldrich) are used for AuNR-dye conjugation. Piranha solution (3:1 solution of \ce{H2SO4} and \ce{H2O2}, $T=100^oC$, $t = 30~mins$), Aqua regia (3:1 mixture of \ce{HCl} and \ce{HNO3}), Acetone, De-ionized water (resistivity $18.2~M\Omega~cm$) were used for cleaning purposes.

\subsection{Synthesis of AuNR}
Gold nanorods (AuNRs) of high monodispersity were synthesized through a seed-mediated growth process\cite{nikoobakht2003preparation,khawas2024directing}. Initially, the seed solution was prepared by combining CTAB solution (0.2 M, 5 ml) with gold solution (0.5 mM, 5 ml), followed by the addition of freshly prepared ice-cold NaBH$_4$ (10 mM, 600 µL) under vigorous stirring at 1000 rpm for 2 minutes. The colour of the seed solution changed to brown, and it was left at room temperature for 30 minutes before utilisation.
Gold solution (5 mM, 9 ml) and silver nitrate solution (0.1 M, 112 µL) were added to CTAB solution (0.2 M) for the growth solution. Subsequently, HCl (1.2 M, 112 µL) was introduced into the mixture. A mild, freshly prepared reducing agent, ascorbic acid (10 mM, 5.5 ml), was added under gentle stirring, resulting in the solution turning colourless. Finally, 75 µL of the seed solution was gently added into the growth solution and stirred gently for 10 s. The solution was left at room temperature for 12 hours to allow the AuNRs to grow.
The resulting growth solution was centrifuged twice at 12000 rpm for 15 minutes to eliminate excess CTAB and other contaminants. The final suspension was stored in a dark place at room temperature for subsequent use.\par
The prepared AuNRs are characterized using UV-Visible spectroscopy, Transmission electron microscopy (TEM) and Zeta potential measurement, all are elaborated in the supplementary section 1.

\subsection{Surface preparation}
Hydrophobic surfaces were prepared by coating \textit{Si} wafers with hydrophobic silane (PFOTS) using vapour deposition method\cite{munief2018silane}. \textit{Si} wafers (p-type, $<100>$) obtained from Centre of Excellence in Nanoelectronics (CEN), IIT Bombay, were cut into $ 0.5~cm \times 0.5~cm $ substrates for ESA experiments. Initially, the surface was cleaned with D.I. water and acetone. After drying under $N_2$, it was washed with freshly prepared Piranha solution (3:1) to remove all impurities, making the surface hydrophilic. The Piranha-cleaned substrates were thoroughly cleaned with D.I. water, dried under $N_2$, and dried in an oven for 2 hours to remove the moisture.  
$10~\mu L$  of PFOTS was pipetted into a petri dish and kept inside a vacuum desiccator along with the substrates. The pressure level inside the desiccator decreased slowly over time, and deposition was carried out for 6 hours. After deposition, the surface was cleaned with acetone, followed by D.I. water, and dried under $N_2$ again. This removes all silane molecules that are physically adsorbed, keeping only the covalently bonded molecules.
Estimations of surface free energy (SFE) and surface roughness (using AFM) for the coated hydrophobic surfaces are given in supplementary section 2.

\subsection{Experimental methods}
In this work, we investigate the formation of ordered self-assembled structures by AuNRs on hydrophobic silicon (\textit{Si}) surfaces using the droplet evaporation technique. The synthesized AuNRs with an aspect ratio  $\approx 3.3$ (Fig.S2) were made into different stock concentrations from $2~nM$ to $30~nM$. The colloidal stability of these AuNRs is maintained through the attachment of the cationic surfactant CTAB via electrostatic interparticle repulsion.
$2~\mu L$ droplets of AuNR were drop cast on the hydrophobic surface and left to evaporate under ambient conditions ($24^oC$, 45\% RH). The droplet evaporation was recorded using a high-speed camera connected to an optical tensiometer (Attention Theta Flex, Biolin Scientific).
The \textit{in situ} data, such as the contact angle ($\theta$), droplet diameter ($D$), height ($H$) and related parameters, are analyzed using the OneAttension software. The software fits the Young-Laplace equation\cite{berry2015measurement} to the shape of the droplet to find these parameters. 
The correlation between the evaporation profiles from the tensiometer and SEM micrographs of dried samples was investigated to elucidate the relationship between particle deposition and evaporation. 

\section{Results and discussion}

\subsection{Evaporation dynamics}
We examine the micro- and nanoscale morphology of anisotropic gold nanorods assembly by depositing aqueous colloidal suspensions onto hydrophobic substrates while tuning particle-substrate interactions. The electrostatic interaction between the positively charged CTAB coating on AuNRs and the negatively charged PFOTS coating on the \textit{Si} substrate favors the adsorption of AuNRs onto the solid surface. The substrate is negatively charged because of highly electronegative fluorine terminated chains. However, the colloidal suspension is dispersed in aqueous media, making the interaction between the droplet suspension and the substrate unfavorable.  The hydrophobicity of the substrate is evident from its surface free energy value ($\gamma_{SV}\sim16.5~mN/m$) and water contact angle ($\theta_w\sim108^o$) (see SI Table S1). \par
\begin{figure*}[h!]
    \includegraphics[width=0.9 \textwidth]{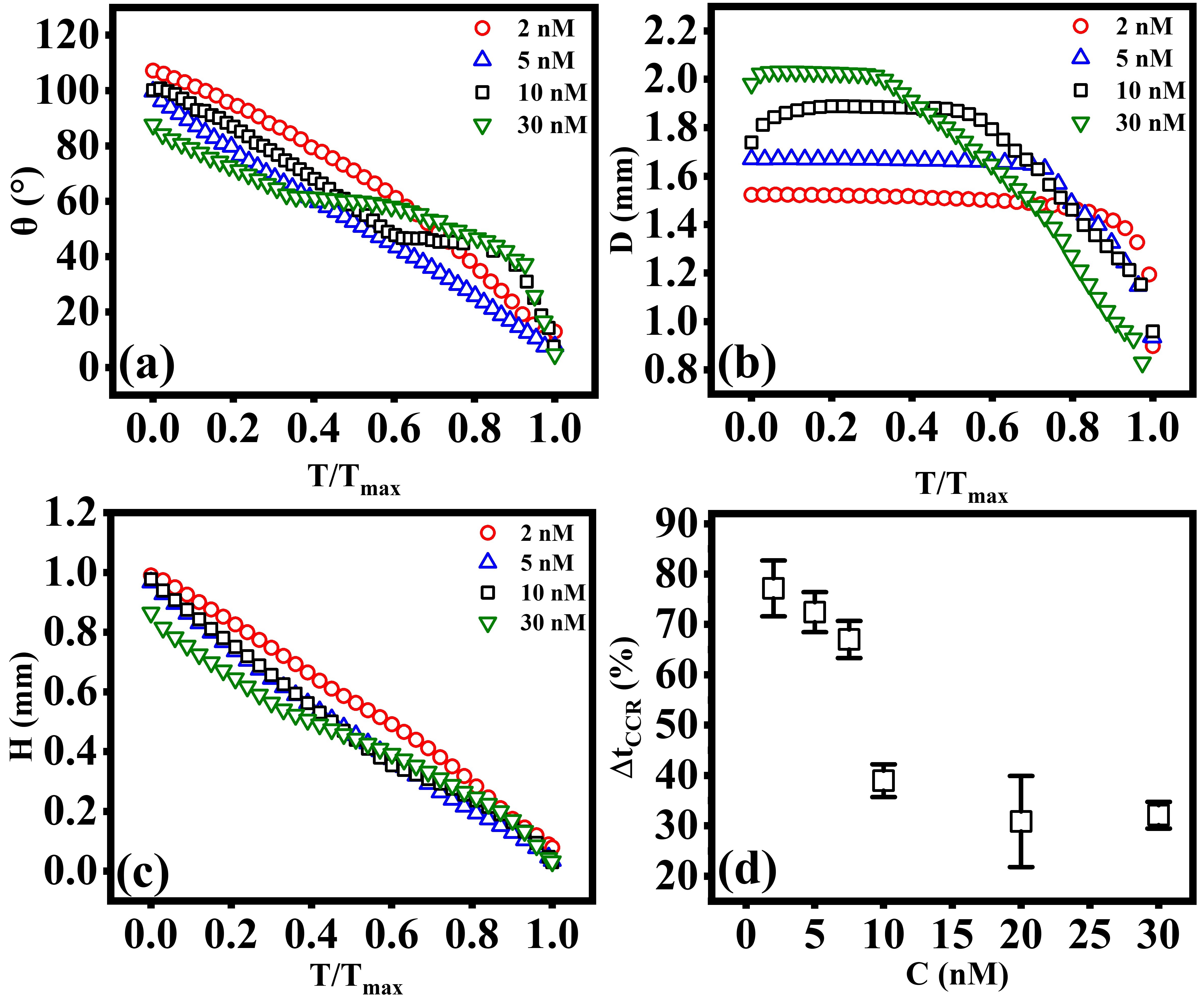}
    \caption{Time evolution of (a) contact angle $\theta$, (b) contact diameter D, and (c) height H of the AuNR colloidal droplet of different concentrations obtained from the optical tensiometer. The percentage of time the droplet evaporated under CCR mode is plotted against nanorod concentration (d)}
    \label{fig:Fig one}
\end{figure*}

The drop-casting of the colloidal suspension onto the hydrophobic substrate results in the formation of semi-spherical droplets with a contact angles ($\theta$) in the range of $110$\textdegree - $85$\textdegree, depending on the concentration of the nanorods in the suspension (Fig.\ref{fig:Fig one}).
The size and shape of a droplet are defined by its $\theta$, $H$ and $D$, which are influenced by the interfacial properties of the solid-liquid ($SV$), solid-liquid ($SL$) and liquid-vapor ($LV$) interfaces.
Equilibrium contact angle ($\theta_0$) is related to surface tension forces along $LV$, $SV$, and $SL$ interfaces by,
\begin{equation}
    \gamma_{LV}cos( \theta_0)=\gamma_{SV}-\gamma_{SL}
\end{equation}\par
where $\gamma_{LV}$, $\gamma_{SV}$, and $\gamma_{SL}$ are the interfacial surface tensions of $LV$, $SV$, and $SL$ interfaces, respectively.

The \textit{in situ} drying profiles obtained from the optical tensiometer corresponding to the droplet's $\theta$, $D$, and $H$ with dimensionless time ($T/T_{max}$) at varying particle concentrations are shown in Fig.\ref{fig:Fig one}(a-c). $T_{max}$ is the total drying time for the corresponding colloidal drop, as recorded by the tensiometer. We report that the evaporation of the colloidal AuNR droplet on the PFOTS-coated Si surface follows distinct drying regimes: (i) short initial spread of the TPCL, (ii) constant contact radius (CCR) mode followed by (iii) constant contact angle (CCA) mode, and finally, (iv) droplet collapse [Fig.\ref{fig:Fig one}(a,b)]. At low concentrations ($C<10~nM$), AuNR droplets exhibit an initial contact angle $\theta>100$\textdegree (Fig.\ref{fig:Fig one}a) and an extended period of evaporation characterized by the CCR mode. Here, the droplet is pinned to the surface and evaporates with a constant contact diameter until the drop collapses, as can be seen in Fig.\ref{fig:Fig one}(b).\par

High-concentration samples ($C\geq 10~nM$) initially exhibited CCR mode, transitioning to CCA mode before final collapse (Fig. \ref{fig:Fig one}a). The onset to CCA mode was found to be dependent on particle concentration, occurring at an earlier stage of evaporation with increasing concentration. Furthermore, we observed that at high-concentrations, the TPCL initially advanced before getting pinned (Fig.\ref{fig:Fig one}b). This advancement of TPCL can be attributed to the initial spreading of droplet facilitated by the adsorption of free surfactant on the $\textit{SV}$ interface\cite{ahmad2019self}.
In the case of a hydrophobic surface, free CTAB in the droplet diffuses at the $\textit{SV}$ interface and adsorbs onto the substrate through hydrophobic bonding (hydrophilic head facing up). This adsorption makes the area near the contact line hydrophilic, resulting in initial spreading, signature of which is seen in the brief advancing of TPCL for high-concentration samples ($10~nM$ and $30~nM$ in Fig.\ref{fig:Fig one}(b)). This initial spreading contrasts the `autophobic effect' reported for hydrophilic surfaces \cite{khawas2023anisotropic}.
TPCL attains stability shortly after the initial spreading through pinning and follows the CCR evaporation mode. For samples with higher concentrations of nanorods ($C\geq10~nM$), the substrate appeared less hydrophobic, as evidenced by a reduction in the initial contact angle and an increment in contact diameter [Fig.\ref{fig:Fig one}(a,b)]. In literature, surfactant-laden droplets on perfluoroalkyl silane-coated surfaces have shown increased wettability as the surfactant concentration has increased \cite{kwiecinski2019evaporation}.

\par
The percentage of time the droplet remained pinned to the substrate ($\Delta$$t_{CCR}$) was calculated by fitting the diameter profile to a straight line and plotted as a function of nanorod concentration (Fig.\ref{fig:Fig one}d).
Drying profiles show a transition from CCR to CCA mode as particle concentration increases, and thus, $\Delta$$t_{CCR}$ follows a sharp decrease between $5~nM$ and $10~nM$. This observation of TPCL receding for $C>5~nM$  range can be associated with the saturation of $LV$ interface by surfactant as well as the anisotropic particles.

Diffusion and adsorption of free surfactant along the droplet interface play a crucial role in controlling the TPCL dynamics.
The movement of CTAB molecules from the bulk liquid towards the \textit{LV} interface is a recognized phenomenon\cite{shao2020role,lee2008kinetics}.
When droplet evaporation occurs in CCR mode, this migration leads them to occupy the \textit{LV} interface, leading to a gradient in interfacial tension ($\Delta\gamma_{LV}$) from the periphery to the apex. The solutal Marangoni convection velocity ($u_M$) is dependent on the variation in interfacial tension ($\Delta\gamma$), as $u_M\propto \Delta\gamma/\mu$, where $\mu$ is the viscosity of the solvent\cite{johansson2022molecular}. The stress developed due to this $\Delta\gamma_{LV}$ exerts a dragging force on the contact line, resulting in depinning.
In case of the low-concentration samples, the total CTAB concentration in the droplet is insufficient to create strong Marangoni stress; thereby, the droplet continues to evaporate in CCR mode. A rough estimation of the number of nanorods required to saturate the hemispherical \textit{LV} interface can be found using $N_{sat}\approx \frac{R^2+H^2}{rl}$, where $r$, $l$ are the radius and length of the rods, respectively. For the $10~nM$ sample, the number of particles within a droplet $2~\mu L$ is $\sim 2.6N_{sat}$, which can readily cover the \textit{LV} interface. For the $5~nM$ sample, the droplet contains $\sim 1.4N_{sat}$ particles, suggesting a lesser particle movement towards \textit{LV} interface owing to a stronger capillary flow. This suggests that the solutes occupy and saturate the $LV$ interface for $C>5~nM$, hence facilitating the transition from CCR to CCA mode of evaporation.
  
Finally, the droplet collapses when the volume gets significantly less, caused by mechanical instabilities.
\par
\subsection{Characteristics of dried particulate deposits}

\begin{figure*}[h!]
    \includegraphics[width=0.9 \textwidth]{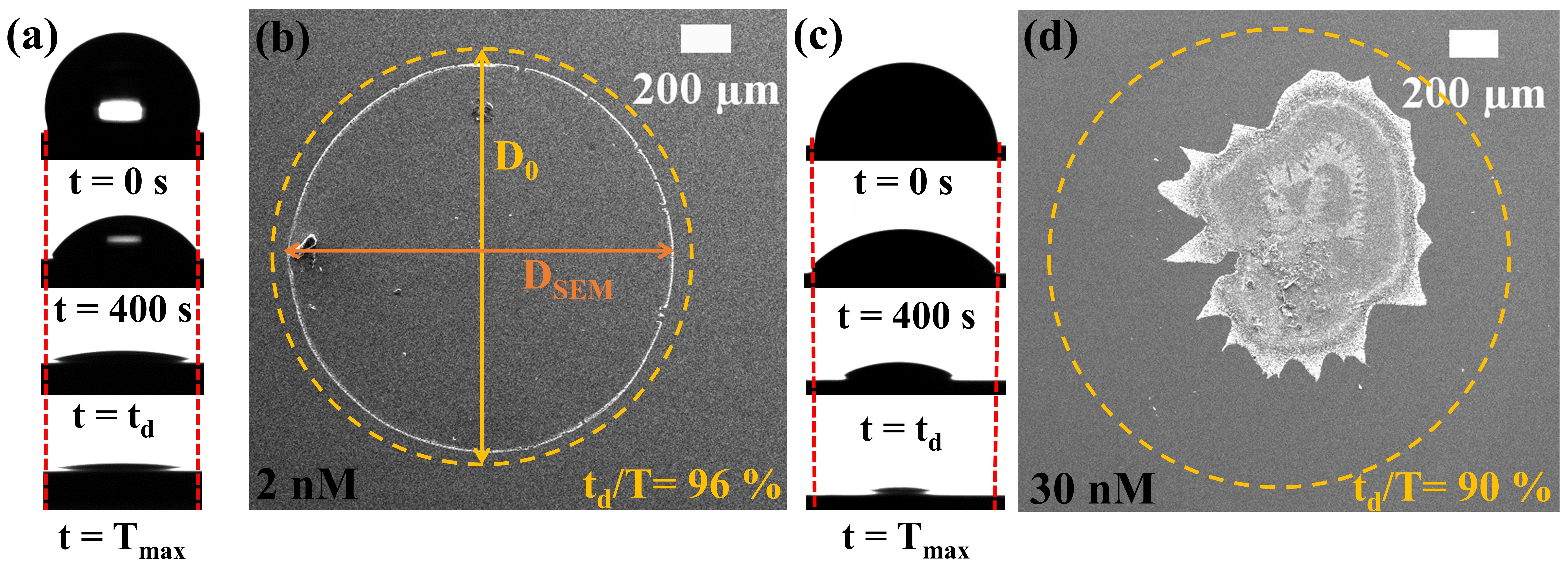}
    \caption{(a,c) Droplet images of $2~nM$ and $30~nM$ AuNR, at different instances from $t=0~s$ to $t=T_{max}$, obtained from the optical tensiometer. $t_d$ represents the onset of deposition, which occurs towards the end of drying. Low-magnification SEM micrographs of $2~nM$ and $30~nM$ (b,d) AuNR dried particulate show coffee-ring to uniform deposition. Dashed yellow circles represent $D_{0}$ (not correct to the scale), giving an idea of the extent of TPCL retraction.}
    \label{fig:Fig two}
\end{figure*}

The surface characteristics of the dried particulate of CTAB-AuNR droplet on hydrophobic \textit{Si} surface and insights into the mechanism of pattern formation were obtained by correlating \textit{in situ} optical tensiometry data with \textit{ex situ} electron and optical microscopy imaging. 
The real-time images of the drying droplet for the lowest ($2~nM$) and highest ($30~nM$) AuNR concentration are shown in Fig.\ref{fig:Fig two}(a,c). The low-magnification SEM micrographs in Fig.\ref{fig:Fig two}(b,d) display the dried surface patterns for these two cases, illustrating the transition in deposit morphology as AuNR concentration increases. The initial droplet diameter ($D_{0}$) was determined from the evaporation profile by tensiometer, while the statistical diameter of the dried deposit ($D_{SEM}$) was calculated from SEM micrographs using ImageJ software \cite{schneider2012nih}. It was observed that the estimated $D_{0}$, represented by a dotted circle, is greater than $D_{SEM}$ [Table S2]. Evaporation time at $D(t)=D_{SEM}$ is estimated by correlating Fig.\ref{fig:Fig one}b and Fig. \ref{fig:Fig two}(b,d) and is attributed as the onset time of nanorod deposition ($t_d$). The droplet images at $t=t_d$ [Fig. \ref{fig:Fig two}(a,c)] indicate that evaporation has progressed to the collapse stage, with AuNR deposition on the hydrophobic surface occurring at late stages of evaporation when $t_d\geq0.9T_{max}$.

It is an interesting feature since the pinning of the TPCL that was observed to occur at the very initial stage of dropcast  (Fig.\ref{fig:Fig one}b), did not lead to coffee-ring formation via the nanorod deposition. Additionally, nanorods are not deposited outside the coffee-ring, as confirmed by the SEM micrographs [Fig.\ref{fig:Fig two}(b,d)], an observation which contrasts with the deposition patterns on a hydrophilic surface\cite{khawas2023anisotropic}. This remains consistent even when the droplet recedes following the initial CCR mode, as demonstrated in the $30~ nM$ system shown in Fig.\ref{fig:Fig two}c. The experimental observations reveal a notable difference for AuNR depositions on surfaces with favorable and non-favorable interactions, indicating a strong influence of nanoscale interactions on macroscale deposition patterns. 
\par
The weak capillary flow inside the droplet carries AuNRs towards the TPCL during the CCR mode. However, the hydrophilic AuNRs are repelled by the hydrophobic surface, preventing particle deposition. Receding TPCL draws the AuNRs along with it until the onset of deposition. 
At $t_d$, the droplet apex height, \textit{H} is significantly small, $0.09 ~mm$ and $0.17~ mm$ for $2~ nM$ and $30 ~nM$, respectively (Fig.\ref{fig:Fig one}c). Consequently, the height at the TPCL is even less.
The observation of particle deposition at a later stage ($t=t_d$) suggests that the droplet height reaches a certain minimum value at this stage, below which the AuNRs are no longer suspended and are compelled to settle onto the substrate. Low-magnification SEM images in Fig.\ref{fig:Fig two}(b,c), reveals a surface morphological transition from a coffee-ring to a uniform deposition as AuNR concentration in droplet suspension increases.
Systematic evolution of the micro-scale surface deposits at varying concentrations is shown in Fig.S5. For $C<10~nM$, we observe a distinct coffee-ring with scattered nanorod deposition at the inner regions. Nanorod clusters deposited at the inner regions of the coffee-ring became more prominent from $C\geq10~nM$.\par

The interplay of capillary and Marangoni flow determines the translation of AuNRs within the droplet, causing the particles to be carried toward or away from the interfaces. In CCR mode, the capillary flow dominates the Marangoni flow, compelling the nanorods to migrate toward the TPCL. However, unfavorable particle-substrate interactions result in the concentration build-up of rods near the TPCL, establishing a concentration gradient relative to the bulk. This gradient in the concentration of nanorods and surfactant molecules, along TPCL induces a strong Marangoni flow, forcing the TPCL to get de-pinned and recede. AuNR clusters are dragged by the receding TPCL to the inner regions of the droplet. The coffee-ring width reveals an non-monotonic dependence on particle concentration as discussed in supplementary section 5. With an increase in AuNR concentration, the Marangoni flow becomes stronger to facilitate nanorod deposition in the inner regions, thereby forming a uniform deposition.
\par 
\begin{figure*}[h!]
    \includegraphics[width=0.9 \textwidth]{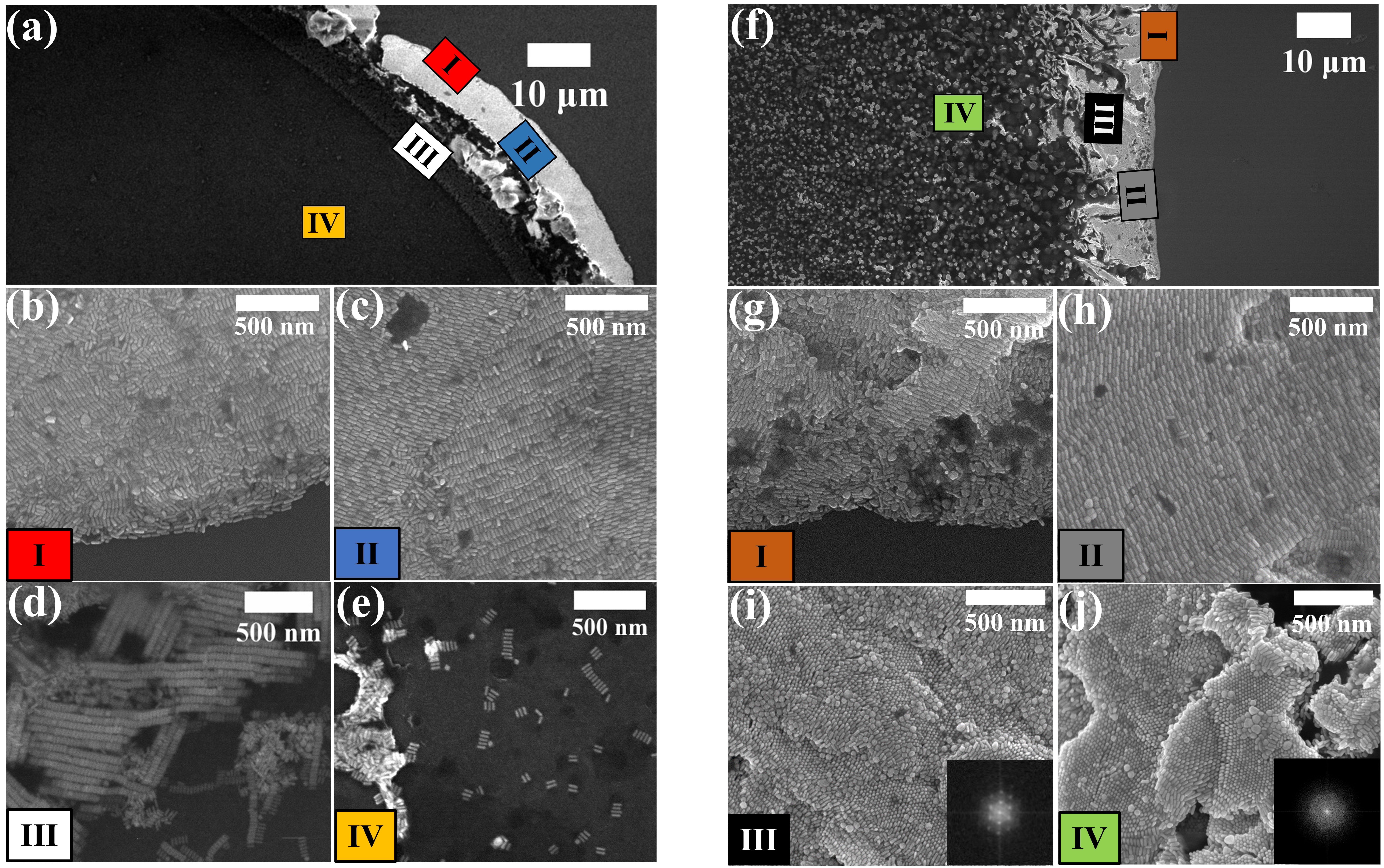}
    \caption{SEM micrographs of dried depositions of AuNR nanocolloid droplet at $2~nM$ (left) and $30~nM$ (right). (a,f) Represents the low-magnification images in the case of $2~nM$ and $30~nM$, respectively. High-magnification SEM micrographs of different regions are marked as $I$, $II$, $III$, and $IV$ for $2 ~nM$(b-e) and $30~nM$(g-j) respectively. The FFT of region $III$ and $IV$ are shown in inset to (i,j), respectively, showing triangular lattice in both regions.}
    \label{fig:Figure 3}
\end{figure*}
We further analyze the nanoscale assembly of AuNRs to understand length-scale-dependent assembly and the influence of microscale features and substrate wettability. 
In Fig. \ref{fig:Figure 3}(a,f), we mark the different assembly regions on the SEM micrographs corresponding to $2~nM$ (left panel) and $30~nM$ (right panel) samples, respectively. It is clear that at low-concentration ($2~nM$), a well-defined, compact coffee-ring is formed whereas, at high concentration ($30~nM$), the presence of Marangoni flow drives the nanorods toward the inner regions, resulting in a coffee-ring with diffused inner boundary.  Fig.\ref{fig:Figure 3}(a-e) and Fig.\ref{fig:Figure 3}(f-j) represents high-magnification SEM micrographs corresponding to $2~nM$ and $30~nM$ of AuNR assembly, respectively. Region $I$ and $II$ represent the outer and middle part of the coffee-ring, region $III$ is the inner vicinity, and region $IV$ is inside the coffee-ring boundary. Region $I$ in low-concentration deposits ($C< 5~nM$) comprises of a line of AuNRs aligned with the contact line, followed by particles arranged isotropically. For a concentration of $30~nM$ (Fig.\ref{fig:Figure 3}g), region $I$ exhibits similar characteristics. Region $II$ displays highly organized smectic domains of AuNRs [Fig.\ref{fig:Figure 3}(c,h)]. The characteristics of regions $I$ and $II$ are consistent across all experimental concentrations. It is important to note that the SEM micrographs represent the average features from the different deposit regions, obtained from multiple experimental repeats. Interestingly contrasting features for nanoscale assembly with increasing AuNR concentration occurs in regions $III$ and $IV$. Region $III$ in the $2~nM$ sample features side-by-side assembly of AuNR forming linear arrays (Fig.\ref{fig:Figure 3}d). These are mostly monolayers, but as concentration increases, they grow into larger clusters. For high-concentration samples ($C\geq10~nM$), AuNR undergoes a smectic to vertical phase transition in region $III$ (Fig.\ref{fig:Figure 3}i). This region consists of highly crystalline, multi-layers of AuNRs arranged into triangular lattices as shown in the Fast Fourier transform (FFT) of the SEM micrograph (inset to Fig.\ref{fig:Figure 3}i). The FFTs shown in the insets of Fig.\ref{fig:Figure 3}(i,j) reveal long-range ordering at region III as compared to region IV due to the presence of defects at the cluster boundaries, also evident in Fig.\ref{fig:Figure 3}f. 
The AuNR clusters formed within the droplet, orient themselves vertically while depositing, likely to maximize the packing and accommodate larger clusters.
In the case of low-concentrations ($C<10~nM$), the inside regions of the deposits consist of scattered AuNRs, whereas at high-concentrations, we observe micron-sized clusters of AuNR [Fig.\ref{fig:Figure 3}(e,j)]. These clusters in region IV at high-concentration constitute numerous island-like depositions, resulting in a transition from the coffee-ring to uniform deposition,  as it appears at the macro-scale.
\par
The observed structural change from smectic to vertical arrays is evident from the comparison of the nanoscale assembly in region III from low and high-concentration samples in Fig.\ref{fig:Figure 3}(d-e) and Fig.\ref{fig:Figure 3}(i-j) respectively. The data indicates that at low-concentration, small clusters of AuNRs are formed in solution where rods are aligned smectically. In the case of high-nanorod concentrations, smectic AuNR cluster grows into three-dimensional clusters, and depositions in vertical arrays are favorable to maximize the cluster density at the surface. In all cases, cluster deposition on the substrate happens at the collapse stage of droplet evaporation due to unfavorable interaction between the suspension and substrate. \par
\begin{figure*}[h!]
    \includegraphics[width={0.9\textwidth}]{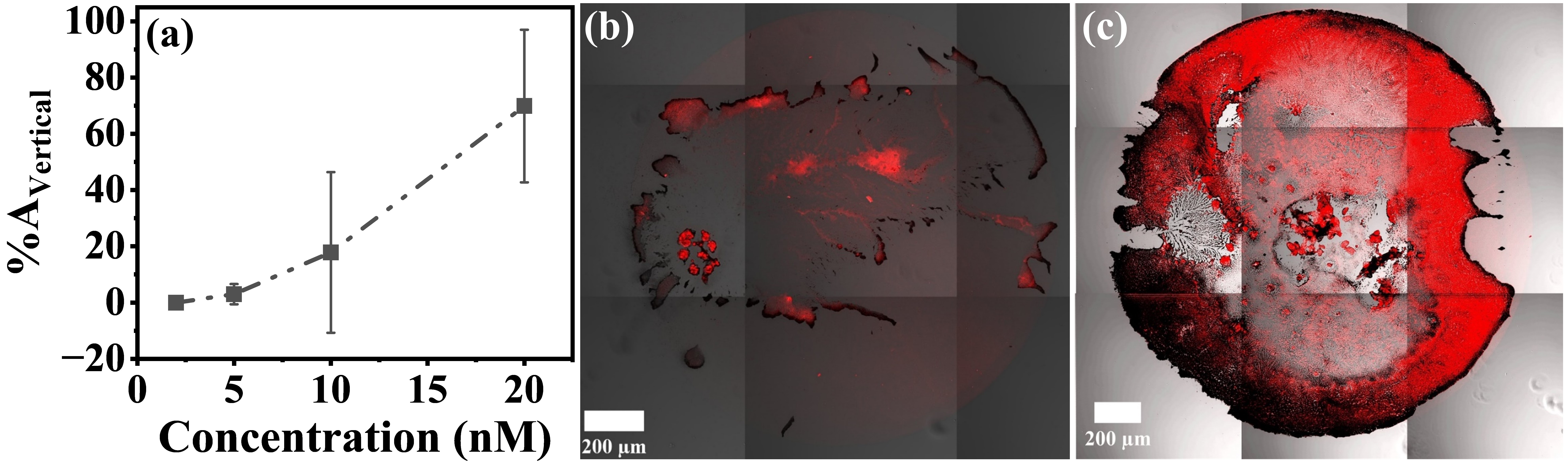}
    \caption{(a) The percentage of total deposit area occupied by vertically arranged AuNRs on hydrophobic substrates for different AuNR concentrations. Surface morphology of dried depositions of Nile red-tagged AuNR drops at concentrations $2~nM$ (b) and $20~nM$ (c) on PFOTS-coated coverslips obtained from confocal laser scanning microscope.}
    \label{fig:Figure 4}
\end{figure*}

The formation of vertically ordered AuNRs clusters in regions $III$ and $IV$, increases with nanorod concentration. Fig.\ref{fig:Figure 4}a shows the percentage of vertical rods ($\%A_{Vertical}$) with varying concentrations. Using high-magnification SEM images we calculated the area occupied by vertically arranged nanorods to the total area. These estimations are repeated for several images for statistical representation. Introducing non-favorable substrate interactions significantly increases the fraction of vertical rods on hydrophobic substrates. A comparative analysis of nano-scale configurations of AuNR on hydrophobic and hydrophilic surfaces unveils a substantially small fraction of vertical nanorods on the hydrophilic substrate, and the majority of the assembly is in isotropic/smectic phases (Fig.S7).

To confirm the presence of Marangoni flow and the resulting high percentage of vertical rods ($\%A_{Vertical}$), on hydrophobic substrates, Nile Red dye-tagged CTAB-AuNRs were evaporated under similar conditions. As Nile Red is a hydrophobic dye that is encapsulated within the CTAB bilayers\cite{mclintock2013stabilized,yue2020hierarchical}, tracking the intensity of the red dye post-deposition can be correlated with the fluid flow behavior within the droplet. Fig.\ref{fig:Figure 4}(b,c) are reconstructed confocal images of the dried droplets of $2~nM$ and $20~nM$ of colloidal AuNR, respectively. While the $2~nM$ sample exhibits negligible red intensity in the inner region of the coffee ring, indicating minimal dye and CTAB deposition, the $20~nM$ sample, in contrast, exhibits significant Nile Red dye and CTAB deposition in regions $III$ and $IV$. These observations, combined with \textit{in situ} evaporation profiles and SEM micrographs, confirm strong Marangoni flow in the high-concentration samples, facilitating vertical AuNR assembly. The observed transition in macroscale deposition pattern from coffee-ring (low-concentration) to uniform deposition (high concentration) can also be attributed to the dominance of Marangoni flow. Based on these measurements and observations, we propose the following mechanism for the surface deposition of AuNR on hydrophobic substrates.
\par

\subsection{Mechanism}
\label{subsec:mech}
The observation that the deposition of ordered nanorod structures on PFOTS-coated \textit{Si} surface occurs at the latter evaporation stage ($ t \geq t_d$) even though the TPCL remains pinned since early stage of droplet evaporation is an important outcome of these experiments.
Further, the macro-scale deposition pattern exhibits a transition from the coffee-ring to uniform deposition pattern with increasing nanoparticle concentration. At the nanoscale, there is a formation of crystalline smectic AuNR within the coffee ring, and the inner region of the coffee ring exhibits the vertical clusters of nanorods.\par

\begin{figure*}[h!]
    \includegraphics[width=0.9 \textwidth]{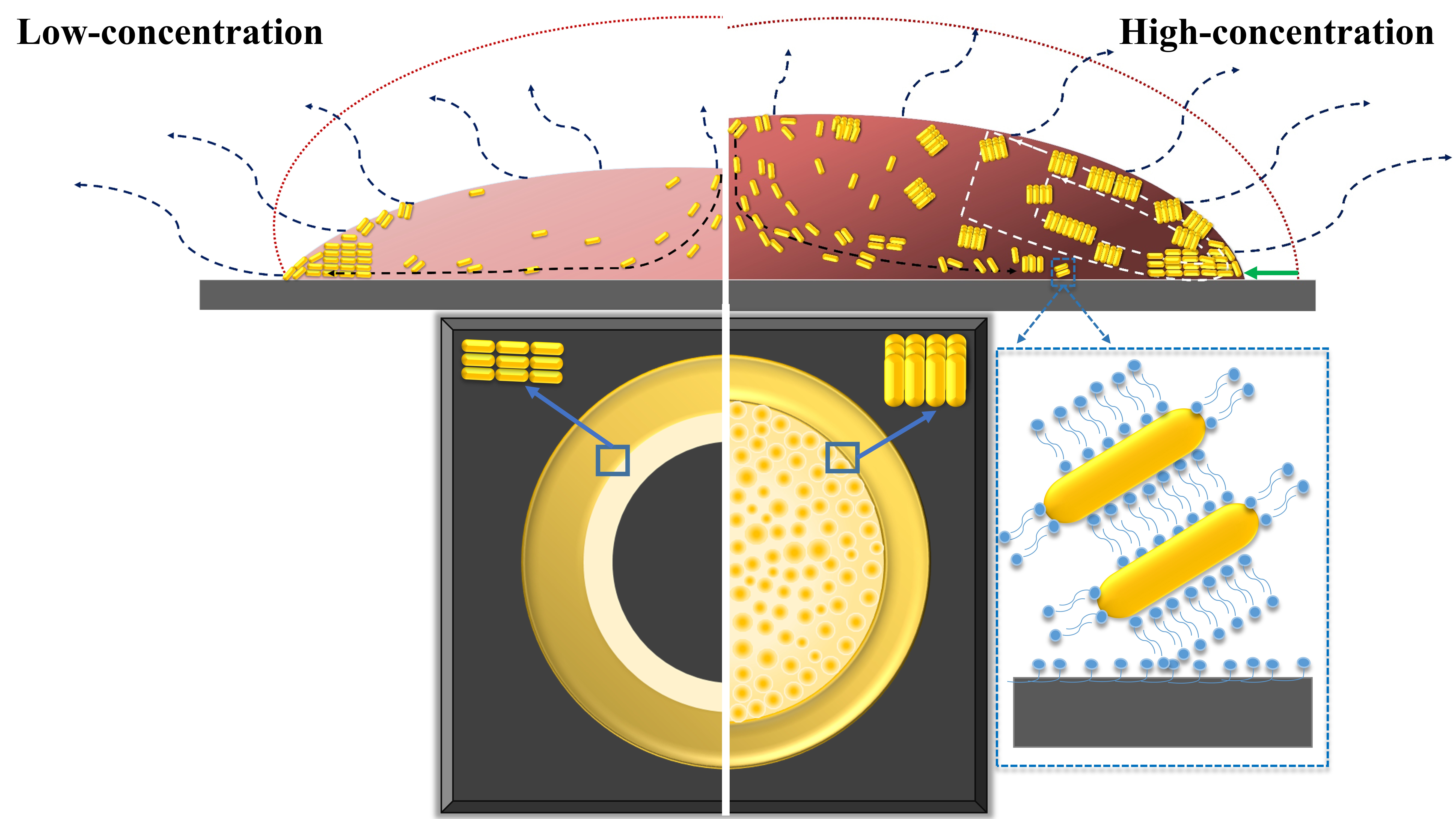}
    \caption{The schematic shows a snapshot of the final stage of evaporation ($t \geq t_d$) of the colloidal AuNR droplet, at low (left) and high (right) concentrations. The red dotted line denotes the initial shape of the droplet ($t=0$). At low concentration, the TPCL remains pinned throughout evaporation, while at high concentration, its recession is shown by a green arrow. Capillary and Marangoni flows are represented by black and white dotted lines, respectively. The macroscale surface morphology is illustrated at the bottom panel. The bottom-right inset illustrates the interaction between CTAB-capped AuNR clusters and the hydrophobic surface, which is significant at high concentrations.}
    \label{fig:figure 5}
\end{figure*}

Initially, the droplet gets pinned to the surface heterogeneities, leading to droplet evaporation under CCR mode. At this stage, capillary flow directs the solutes toward the contact line, leading to an increase in the concentration of nanorods near the TPCL. However, unfavorable interactions between hydrophillic AuNRs and the hydrophobic substrate prevent deposition at this point. A meta-stable energy minimum exists between two interacting AuNRs , forming ordered nanorods arrays in the colloidal solution \cite{xie2013controllable}. This means that the AuNRs will form clusters inside the droplet, and as the drop evaporates, the clusters will grow in size. The inhibition of nanorod deposition on the substrate favors the cluster growth at the \textit{LV} interface forming large domains. Capillary flow gets stronger as the evaporation proceeds (geometry of the droplet becomes less spherical), drawing more AuNRs near the TPCL.  These clusters get deposited at TPCL as coffee-ring, only at the later stage of evaporation when enough solvent loss has happened to reach a significant reduction in droplet's height. This schematic demonstration of the mechanism for AuNR deposition at low and high concentration is presented in Fig.\ref{fig:figure 5}.

As the nanorod concentration increases, droplet pinning duration decreases, and the contact line starts to recede. However, regardless of the mode of evaporation, we confirm that the nanorod deposition occurs only near the droplet's collapse. The sharing of CTAB bilayer between two AuNRs is a recognized process which is an energetically stable configuration\cite{khawas2024directing}. This leads to an increase in the concentration of expelled CTAB in the droplet and enhances solutal gradient along the \textit{LV} interface, $\Delta\gamma_{LV}$, which in turn causes the Marangoni flow. The interplay of capillary and Marangoni flow causes the macro-scale deposition pattern from coffee-ring at low concentration to uniform deposition at high concentration. In the case of low-concentration samples, the particle concentration at the \textit{LV} interface is not sufficient to induce Marangoni flow, resulting in coffee-ring formation (Fig.\ref{fig:Figure 4}). At high concentration the Marangoni flow promotes redistribution of AuNR clusters towards the central regions of the droplet. The island-like deposition of AuNRs at high-concentrations is the result of strong Marangoni flow formed due to the preferential adsorption of CTAB-AuNR at the \textit{LV} interface compared to \textit{SL} interface.\par
The formation of smectic ordered phases at low concentration [Fig.\ref{fig:Figure 3}(d,e)] and vertical arrays of nanorods [region III, Fig.\ref{fig:Figure 3}(i,k)] at high concentration within the coffee-ring deposit  can be explained by considering the particle-particle and particle-substrate interactions at the nanoscale.
At low concentrations, an electrostatic attraction between the substrate and AuNR favors horizontal orientation to the substrate. This leads to the smectic arrangement of AuNR within the coffee-ring as shown in Fig.\ref{fig:Figure 3}d and Fig.\ref{fig:figure 5}. With the increase in concentration, the most probable arrangement between two AuNR involves sharing a CTAB bilayer \cite{khawas2024directing}, causing excess pair of CTAB molecules to be pushed out into the surrounding liquid phase  due to depletion interaction. In the case of a hydrophobic surface, the released surfactant is more prone to adsorb at the \textit{SL} interface\cite{kwiecinski2019evaporation,staniscia2022tuning}.
CTAB adsorption on the \textit{SL} interface occurs such that its hydrophilic head faces away from the substrate (Fig.\ref{fig:figure 5}), decreasing the effective negative-surface potential of the substrate\cite{ahmad2019self}. Further, the formation of such a self-assembled monolayer of CTAB diminishes the hydrophobic forces and induces electrostatic repulsion between the particles and the substrate, favoring the vertical arrangement of AuNRs (Fig.\ref{fig:Figure 3}i). When AuNR concentrations are high, the deposition area at $t=t_d$ decreases due to the receded contact line compared to a low-concentration system. In smaller areas, nanorod arrays tend to align vertically over the substrate, because this arrangement is more energetically favorable and geometrically stable. Due to the geometrical effects of the droplet near TPCL, AuNRs cannot “stand” at the outer edge and are only arranged horizontally (region I and II). This explains why vertical arrangements are negligible near the outer regions of the coffee-ring [Fig.\ref{fig:Figure 3}(g-h)]. At low concentrations, the CTAB available in droplets is insufficient to form a carpet on the hydrophobic film, and hence the attraction of particles and substrates dominates and promotes the smectic alignment of nanorods [Fig.\ref{fig:Figure 3}(d-e)] as shown in Fig.\ref{fig:figure 5}. The experimental results presented here provide insights into the strong influence of nanoscale interactions on macroscale deposition patterns. These studies are important for several applications that exploit colloidal suspension for surface coating.

\section{Conclusion}
In this study, we analyzed the dried patterns formed by colloidal droplets of CTAB-capped AuNRs of varying concentrations, with a specific focus on the influence of non-favorable particle-substrate interactions. Our results unravel a strong correlation between the \textit{in situ} drying profiles, measured using an optical tensiometer, and the deposit pattern, imaged \textit{ex situ} using electron microscopy. An increase in AuNR concentration induces a transition from a coffee-ring to a uniform deposition pattern at the macroscale. At the nanoscale, we report the formation of well-ordered phases. For lower concentrations, AuNRs align smectically but as the concentration increases, the AuNRs self-assemble into vertically aligned triangular lattice structures.\par
Our findings provide valuable insights into the factors controlling the self-assembly of anisotropic nanoparticles. At low-concentrations ($\leq5~nM$), droplets evaporated via CCR mode only, whereas drying of high-concentration droplets proceeded via initial CCR followed by CCA mode. Analysis of drying profiles and deposited patterns suggest that the nanorods get deposited only at the end stages of evaporation ($t_d\geq0.9T_{max}$). The distinct particle deposition we report is a result of preferential adsorption of nanorods between \textit{LV} and \textit{SL} interface, leading to AuNR cluster formation inside the droplet. As the concentration increases, the size of AuNR clusters within the inner vicinity of the coffee-ring grows, resulting in a transition from smectic arrays of AuNRs at low concentrations to vertically aligned rod clusters at high concentrations. We demonstrate that the interplay of capillary and Marangoni flow, which primarily drives fluid flow at the macroscale, significantly impacts the assembly phenomena observed at the nanoscale. By mapping out the conditions leading to different structural arrangements, we aim to establish a framework for designing and producing materials with tailored properties for future optoelectronic applications.

\section*{Conflict of interest}
The authors declare no conflict of interest.

\section*{Acknowledgement}
NPV thanks financial support from the Physics department, IIT Bombay. 
SB acknowledges the financial support from CRNTS, IIT Bombay.
SS acknowledges support from the CRS-UGC-DAE, India.
We thank SAIF IIT Bombay for the FEG-SEM facility, IRCC central facilities for FEG-SEM (Department of Chemical Engineering), Confocal microscope and AFM (Department of Biosciences and Bioengineering).

\begin{suppinfo}
Characterizations for AuNRs, PFOTS-Si substrate, AuNR-Nile red conjugation, \textit{ex situ} SEM micrographs, analysis for CRW and area analysis of vertically oriented AuNRs are provided here.
\end{suppinfo}

\bibliography{references}

\providecommand{\latin}[1]{#1}
\makeatletter
\providecommand{\doi}
  {\begingroup\let\do\@makeother\dospecials
  \catcode`\{=1 \catcode`\}=2 \doi@aux}
\providecommand{\doi@aux}[1]{\endgroup\texttt{#1}}
\makeatother
\providecommand*\mcitethebibliography{\thebibliography}
\csname @ifundefined\endcsname{endmcitethebibliography}  {\let\endmcitethebibliography\endthebibliography}{}
\begin{mcitethebibliography}{52}
\providecommand*\natexlab[1]{#1}
\providecommand*\mciteSetBstSublistMode[1]{}
\providecommand*\mciteSetBstMaxWidthForm[2]{}
\providecommand*\mciteBstWouldAddEndPuncttrue
  {\def\EndOfBibitem{\unskip.}}
\providecommand*\mciteBstWouldAddEndPunctfalse
  {\let\EndOfBibitem\relax}
\providecommand*\mciteSetBstMidEndSepPunct[3]{}
\providecommand*\mciteSetBstSublistLabelBeginEnd[3]{}
\providecommand*\EndOfBibitem{}
\mciteSetBstSublistMode{f}
\mciteSetBstMaxWidthForm{subitem}{(\alph{mcitesubitemcount})}
\mciteSetBstSublistLabelBeginEnd
  {\mcitemaxwidthsubitemform\space}
  {\relax}
  {\relax}

\bibitem[Tang \latin{et~al.}(2017)Tang, Li, Huang, Li, Guo, Luo, Wang, Chu, Li, and Yu]{tang2017efficient}
Tang,~S.; Li,~Y.; Huang,~H.; Li,~P.; Guo,~Z.; Luo,~Q.; Wang,~Z.; Chu,~P.~K.; Li,~J.; Yu,~X.-F. Efficient enrichment and self-assembly of hybrid nanoparticles into removable and magnetic SERS substrates for sensitive detection of environmental pollutants. \emph{ACS applied materials \& interfaces} \textbf{2017}, \emph{9}, 7472--7480\relax
\mciteBstWouldAddEndPuncttrue
\mciteSetBstMidEndSepPunct{\mcitedefaultmidpunct}
{\mcitedefaultendpunct}{\mcitedefaultseppunct}\relax
\EndOfBibitem
\bibitem[Lohse(2022)]{lohse2022fundamental}
Lohse,~D. Fundamental fluid dynamics challenges in inkjet printing. \emph{Annual review of fluid mechanics} \textbf{2022}, \emph{54}, 349--382\relax
\mciteBstWouldAddEndPuncttrue
\mciteSetBstMidEndSepPunct{\mcitedefaultmidpunct}
{\mcitedefaultendpunct}{\mcitedefaultseppunct}\relax
\EndOfBibitem
\bibitem[Vialetto \latin{et~al.}(2024)Vialetto, Gaichies, Rudiuk, Morel, and Baigl]{vialetto2024versatile}
Vialetto,~J.; Gaichies,~T.; Rudiuk,~S.; Morel,~M.; Baigl,~D. Versatile Deposition of Complex Colloidal Assemblies from the Evaporation of Hanging Drops. \emph{Advanced Science} \textbf{2024}, \emph{11}, 2307893\relax
\mciteBstWouldAddEndPuncttrue
\mciteSetBstMidEndSepPunct{\mcitedefaultmidpunct}
{\mcitedefaultendpunct}{\mcitedefaultseppunct}\relax
\EndOfBibitem
\bibitem[Bhardwaj and Agrawal(2020)Bhardwaj, and Agrawal]{bhardwaj2020likelihood}
Bhardwaj,~R.; Agrawal,~A. Likelihood of survival of coronavirus in a respiratory droplet deposited on a solid surface. \emph{Physics of Fluids} \textbf{2020}, \emph{32}\relax
\mciteBstWouldAddEndPuncttrue
\mciteSetBstMidEndSepPunct{\mcitedefaultmidpunct}
{\mcitedefaultendpunct}{\mcitedefaultseppunct}\relax
\EndOfBibitem
\bibitem[Apte \latin{et~al.}(2015)Apte, Joshi, Bhaskar, Joag, and Kulkarni]{apte2015vertically}
Apte,~A.; Joshi,~P.; Bhaskar,~P.; Joag,~D.; Kulkarni,~S. Vertically aligned self-assembled gold nanorods as low turn-on, stable field emitters. \emph{Applied Surface Science} \textbf{2015}, \emph{355}, 978--983\relax
\mciteBstWouldAddEndPuncttrue
\mciteSetBstMidEndSepPunct{\mcitedefaultmidpunct}
{\mcitedefaultendpunct}{\mcitedefaultseppunct}\relax
\EndOfBibitem
\bibitem[Deegan \latin{et~al.}(1997)Deegan, Bakajin, Dupont, Huber, Nagel, and Witten]{deegan1997capillary}
Deegan,~R.~D.; Bakajin,~O.; Dupont,~T.~F.; Huber,~G.; Nagel,~S.~R.; Witten,~T.~A. Capillary flow as the cause of ring stains from dried liquid drops. \emph{Nature} \textbf{1997}, \emph{389}, 827--829\relax
\mciteBstWouldAddEndPuncttrue
\mciteSetBstMidEndSepPunct{\mcitedefaultmidpunct}
{\mcitedefaultendpunct}{\mcitedefaultseppunct}\relax
\EndOfBibitem
\bibitem[Yunker \latin{et~al.}(2011)Yunker, Still, Lohr, and Yodh]{yunker2011suppression}
Yunker,~P.~J.; Still,~T.; Lohr,~M.~A.; Yodh,~A. Suppression of the coffee-ring effect by shape-dependent capillary interactions. \emph{nature} \textbf{2011}, \emph{476}, 308--311\relax
\mciteBstWouldAddEndPuncttrue
\mciteSetBstMidEndSepPunct{\mcitedefaultmidpunct}
{\mcitedefaultendpunct}{\mcitedefaultseppunct}\relax
\EndOfBibitem
\bibitem[Mampallil and Eral(2018)Mampallil, and Eral]{mampallil2018review}
Mampallil,~D.; Eral,~H.~B. A review on suppression and utilization of the coffee-ring effect. \emph{Advances in colloid and interface science} \textbf{2018}, \emph{252}, 38--54\relax
\mciteBstWouldAddEndPuncttrue
\mciteSetBstMidEndSepPunct{\mcitedefaultmidpunct}
{\mcitedefaultendpunct}{\mcitedefaultseppunct}\relax
\EndOfBibitem
\bibitem[Anyfantakis and Baigl(2015)Anyfantakis, and Baigl]{anyfantakis2015manipulating}
Anyfantakis,~M.; Baigl,~D. Manipulating the coffee-ring effect: interactions at work. \emph{ChemPhysChem} \textbf{2015}, \emph{16}, 2726--2734\relax
\mciteBstWouldAddEndPuncttrue
\mciteSetBstMidEndSepPunct{\mcitedefaultmidpunct}
{\mcitedefaultendpunct}{\mcitedefaultseppunct}\relax
\EndOfBibitem
\bibitem[Bhattacharjee and Srivastava(2023)Bhattacharjee, and Srivastava]{bhattacharjee2023ordered}
Bhattacharjee,~S.; Srivastava,~S. Ordered stripes to crack patterns in dried particulates of DNA-coated gold colloids via modulating nanoparticle--substrate interactions. \emph{Soft Matter} \textbf{2023}, \emph{19}, 2265--2274\relax
\mciteBstWouldAddEndPuncttrue
\mciteSetBstMidEndSepPunct{\mcitedefaultmidpunct}
{\mcitedefaultendpunct}{\mcitedefaultseppunct}\relax
\EndOfBibitem
\bibitem[Li \latin{et~al.}(2016)Li, Yang, Li, and Song]{li2016rate}
Li,~Y.; Yang,~Q.; Li,~M.; Song,~Y. Rate-dependent interface capture beyond the coffee-ring effect. \emph{Scientific reports} \textbf{2016}, \emph{6}, 24628\relax
\mciteBstWouldAddEndPuncttrue
\mciteSetBstMidEndSepPunct{\mcitedefaultmidpunct}
{\mcitedefaultendpunct}{\mcitedefaultseppunct}\relax
\EndOfBibitem
\bibitem[Srivastava \latin{et~al.}(2020)Srivastava, Wahith, Gang, Colosqui, and Bhatia]{srivastava2020dual}
Srivastava,~S.; Wahith,~Z.~A.; Gang,~O.; Colosqui,~C.~E.; Bhatia,~S.~R. Dual-Scale Nanostructures via evaporative assembly. \emph{Advanced Materials Interfaces} \textbf{2020}, \emph{7}, 1901954\relax
\mciteBstWouldAddEndPuncttrue
\mciteSetBstMidEndSepPunct{\mcitedefaultmidpunct}
{\mcitedefaultendpunct}{\mcitedefaultseppunct}\relax
\EndOfBibitem
\bibitem[Deegan(2000)]{deegan2000pattern}
Deegan,~R.~D. Pattern formation in drying drops. \emph{Physical review E} \textbf{2000}, \emph{61}, 475\relax
\mciteBstWouldAddEndPuncttrue
\mciteSetBstMidEndSepPunct{\mcitedefaultmidpunct}
{\mcitedefaultendpunct}{\mcitedefaultseppunct}\relax
\EndOfBibitem
\bibitem[Ristenpart \latin{et~al.}(2007)Ristenpart, Kim, Domingues, Wan, and Stone]{ristenpart2007influence}
Ristenpart,~W.; Kim,~P.; Domingues,~C.; Wan,~J.; Stone,~H.~A. Influence of substrate conductivity on circulation reversal in evaporating drops. \emph{Physical review letters} \textbf{2007}, \emph{99}, 234502\relax
\mciteBstWouldAddEndPuncttrue
\mciteSetBstMidEndSepPunct{\mcitedefaultmidpunct}
{\mcitedefaultendpunct}{\mcitedefaultseppunct}\relax
\EndOfBibitem
\bibitem[Nguyen and Stebe(2002)Nguyen, and Stebe]{nguyen2002patterning}
Nguyen,~V.~X.; Stebe,~K.~J. Patterning of small particles by a surfactant-enhanced Marangoni-B{\'e}nard instability. \emph{Physical Review Letters} \textbf{2002}, \emph{88}, 164501\relax
\mciteBstWouldAddEndPuncttrue
\mciteSetBstMidEndSepPunct{\mcitedefaultmidpunct}
{\mcitedefaultendpunct}{\mcitedefaultseppunct}\relax
\EndOfBibitem
\bibitem[Gelderblom \latin{et~al.}(2022)Gelderblom, Diddens, and Marin]{gelderblom2022evaporation}
Gelderblom,~H.; Diddens,~C.; Marin,~A. Evaporation-driven liquid flow in sessile droplets. \emph{Soft Matter} \textbf{2022}, \emph{18}, 8535--8553\relax
\mciteBstWouldAddEndPuncttrue
\mciteSetBstMidEndSepPunct{\mcitedefaultmidpunct}
{\mcitedefaultendpunct}{\mcitedefaultseppunct}\relax
\EndOfBibitem
\bibitem[Li \latin{et~al.}(2016)Li, Li, Zhou, Tang, Yu, Xiao, Wu, Xiao, Zhao, Wang, \latin{et~al.} others]{li2016evaporative}
Li,~P.; Li,~Y.; Zhou,~Z.-K.; Tang,~S.; Yu,~X.-F.; Xiao,~S.; Wu,~Z.; Xiao,~Q.; Zhao,~Y.; Wang,~H.; others Evaporative self-assembly of gold nanorods into macroscopic 3D plasmonic superlattice arrays. \emph{Adv. Mater} \textbf{2016}, \emph{28}, 2511--2517\relax
\mciteBstWouldAddEndPuncttrue
\mciteSetBstMidEndSepPunct{\mcitedefaultmidpunct}
{\mcitedefaultendpunct}{\mcitedefaultseppunct}\relax
\EndOfBibitem
\bibitem[Anyfantakis \latin{et~al.}(2015)Anyfantakis, Geng, Morel, Rudiuk, and Baigl]{anyfantakis2015modulation}
Anyfantakis,~M.; Geng,~Z.; Morel,~M.; Rudiuk,~S.; Baigl,~D. Modulation of the coffee-ring effect in particle/surfactant mixtures: the importance of particle--interface interactions. \emph{Langmuir} \textbf{2015}, \emph{31}, 4113--4120\relax
\mciteBstWouldAddEndPuncttrue
\mciteSetBstMidEndSepPunct{\mcitedefaultmidpunct}
{\mcitedefaultendpunct}{\mcitedefaultseppunct}\relax
\EndOfBibitem
\bibitem[Khawas and Srivastava(2023)Khawas, and Srivastava]{khawas2023anisotropic}
Khawas,~S.; Srivastava,~S. Anisotropic nanocluster arrays to a diminished zone: different regimes of surface deposition of gold nanocolloids. \emph{Soft Matter} \textbf{2023}, \emph{19}, 3580--3589\relax
\mciteBstWouldAddEndPuncttrue
\mciteSetBstMidEndSepPunct{\mcitedefaultmidpunct}
{\mcitedefaultendpunct}{\mcitedefaultseppunct}\relax
\EndOfBibitem
\bibitem[bri(2020)]{bridonneau2020self}
Self-assembly of nanoparticles from evaporating sessile droplets: Fresh look into the role of particle/substrate interaction. \emph{Langmuir} \textbf{2020}, \emph{36}, 11411--11421\relax
\mciteBstWouldAddEndPuncttrue
\mciteSetBstMidEndSepPunct{\mcitedefaultmidpunct}
{\mcitedefaultendpunct}{\mcitedefaultseppunct}\relax
\EndOfBibitem
\bibitem[Orejon \latin{et~al.}(2011)Orejon, Sefiane, and Shanahan]{orejon2011stick}
Orejon,~D.; Sefiane,~K.; Shanahan,~M.~E. Stick--slip of evaporating droplets: substrate hydrophobicity and nanoparticle concentration. \emph{Langmuir} \textbf{2011}, \emph{27}, 12834--12843\relax
\mciteBstWouldAddEndPuncttrue
\mciteSetBstMidEndSepPunct{\mcitedefaultmidpunct}
{\mcitedefaultendpunct}{\mcitedefaultseppunct}\relax
\EndOfBibitem
\bibitem[Mampallil \latin{et~al.}(2012)Mampallil, Eral, Van Den~Ende, and Mugele]{mampallil2012control}
Mampallil,~D.; Eral,~H.; Van Den~Ende,~D.; Mugele,~F. Control of evaporating complex fluids through electrowetting. \emph{Soft Matter} \textbf{2012}, \emph{8}, 10614--10617\relax
\mciteBstWouldAddEndPuncttrue
\mciteSetBstMidEndSepPunct{\mcitedefaultmidpunct}
{\mcitedefaultendpunct}{\mcitedefaultseppunct}\relax
\EndOfBibitem
\bibitem[Mar{\'\i}n \latin{et~al.}(2012)Mar{\'\i}n, Gelderblom, Susarrey-Arce, van Houselt, Lefferts, Gardeniers, Lohse, and Snoeijer]{marin2012building}
Mar{\'\i}n,~{\'A}.~G.; Gelderblom,~H.; Susarrey-Arce,~A.; van Houselt,~A.; Lefferts,~L.; Gardeniers,~J.~G.; Lohse,~D.; Snoeijer,~J.~H. Building microscopic soccer balls with evaporating colloidal fakir drops. \emph{Proceedings of the National Academy of Sciences} \textbf{2012}, \emph{109}, 16455--16458\relax
\mciteBstWouldAddEndPuncttrue
\mciteSetBstMidEndSepPunct{\mcitedefaultmidpunct}
{\mcitedefaultendpunct}{\mcitedefaultseppunct}\relax
\EndOfBibitem
\bibitem[Seyfert \latin{et~al.}(2021)Seyfert, Berenschot, Tas, Susarrey-Arce, and Marin]{seyfert2021evaporation}
Seyfert,~C.; Berenschot,~E.~J.; Tas,~N.~R.; Susarrey-Arce,~A.; Marin,~A. Evaporation-driven colloidal cluster assembly using droplets on superhydrophobic fractal-like structures. \emph{Soft Matter} \textbf{2021}, \emph{17}, 506--515\relax
\mciteBstWouldAddEndPuncttrue
\mciteSetBstMidEndSepPunct{\mcitedefaultmidpunct}
{\mcitedefaultendpunct}{\mcitedefaultseppunct}\relax
\EndOfBibitem
\bibitem[Zheng \latin{et~al.}(2021)Zheng, Cheng, Zhang, Bai, Ai, Shao, and Wang]{zheng2021gold}
Zheng,~J.; Cheng,~X.; Zhang,~H.; Bai,~X.; Ai,~R.; Shao,~L.; Wang,~J. Gold nanorods: the most versatile plasmonic nanoparticles. \emph{Chemical Reviews} \textbf{2021}, \emph{121}, 13342--13453\relax
\mciteBstWouldAddEndPuncttrue
\mciteSetBstMidEndSepPunct{\mcitedefaultmidpunct}
{\mcitedefaultendpunct}{\mcitedefaultseppunct}\relax
\EndOfBibitem
\bibitem[Ming \latin{et~al.}(2008)Ming, Kou, Chen, Wang, Tam, Cheah, Chen, and Wang]{ming2008ordered}
Ming,~T.; Kou,~X.; Chen,~H.; Wang,~T.; Tam,~H.-L.; Cheah,~K.-W.; Chen,~J.-Y.; Wang,~J. Ordered gold nanostructure assemblies formed by droplet evaporation. \emph{Angewandte Chemie International Edition} \textbf{2008}, \emph{47}, 9685--9690\relax
\mciteBstWouldAddEndPuncttrue
\mciteSetBstMidEndSepPunct{\mcitedefaultmidpunct}
{\mcitedefaultendpunct}{\mcitedefaultseppunct}\relax
\EndOfBibitem
\bibitem[Umadevi \latin{et~al.}(2013)Umadevi, Feng, and Hegmann]{umadevi2013large}
Umadevi,~S.; Feng,~X.; Hegmann,~T. Large Area Self-Assembly of Nematic Liquid-Crystal-Functionalized Gold Nanorods. \emph{Advanced Functional Materials} \textbf{2013}, \emph{23}, 1393--1403\relax
\mciteBstWouldAddEndPuncttrue
\mciteSetBstMidEndSepPunct{\mcitedefaultmidpunct}
{\mcitedefaultendpunct}{\mcitedefaultseppunct}\relax
\EndOfBibitem
\bibitem[Liu \latin{et~al.}(2010)Liu, Cui, Gardner, Li, He, and Smalyukh]{liu2010self}
Liu,~Q.; Cui,~Y.; Gardner,~D.; Li,~X.; He,~S.; Smalyukh,~I.~I. Self-alignment of plasmonic gold nanorods in reconfigurable anisotropic fluids for tunable bulk metamaterial applications. \emph{Nano letters} \textbf{2010}, \emph{10}, 1347--1353\relax
\mciteBstWouldAddEndPuncttrue
\mciteSetBstMidEndSepPunct{\mcitedefaultmidpunct}
{\mcitedefaultendpunct}{\mcitedefaultseppunct}\relax
\EndOfBibitem
\bibitem[Thai \latin{et~al.}(2012)Thai, Zheng, Ng, Mudie, Altissimo, and Bach]{thai2012self}
Thai,~T.; Zheng,~Y.; Ng,~S.~H.; Mudie,~S.; Altissimo,~M.; Bach,~U. Self-assembly of vertically aligned gold nanorod arrays on patterned substrates. \emph{Angewandte Chemie International Edition} \textbf{2012}, \emph{35}, 8732--8735\relax
\mciteBstWouldAddEndPuncttrue
\mciteSetBstMidEndSepPunct{\mcitedefaultmidpunct}
{\mcitedefaultendpunct}{\mcitedefaultseppunct}\relax
\EndOfBibitem
\bibitem[Peng \latin{et~al.}(2013)Peng, Li, Li, Dodson, Zhang, Zhang, Lee, Demir, Yi~Ling, and Xiong]{peng2013vertically}
Peng,~B.; Li,~G.; Li,~D.; Dodson,~S.; Zhang,~Q.; Zhang,~J.; Lee,~Y.~H.; Demir,~H.~V.; Yi~Ling,~X.; Xiong,~Q. Vertically aligned gold nanorod monolayer on arbitrary substrates: self-assembly and femtomolar detection of food contaminants. \emph{Acs Nano} \textbf{2013}, \emph{7}, 5993--6000\relax
\mciteBstWouldAddEndPuncttrue
\mciteSetBstMidEndSepPunct{\mcitedefaultmidpunct}
{\mcitedefaultendpunct}{\mcitedefaultseppunct}\relax
\EndOfBibitem
\bibitem[Loeb \latin{et~al.}(2019)Loeb, Kim, Jiang, Early, Wei, Li, and Kim]{loeb2019nanoparticle}
Loeb,~S.~K.; Kim,~J.; Jiang,~C.; Early,~L.~S.; Wei,~H.; Li,~Q.; Kim,~J.-H. Nanoparticle enhanced interfacial solar photothermal water disinfection demonstrated in 3-D printed flow-through reactors. \emph{Environmental science \& technology} \textbf{2019}, \emph{53}, 7621--7631\relax
\mciteBstWouldAddEndPuncttrue
\mciteSetBstMidEndSepPunct{\mcitedefaultmidpunct}
{\mcitedefaultendpunct}{\mcitedefaultseppunct}\relax
\EndOfBibitem
\bibitem[Zijlstra \latin{et~al.}(2009)Zijlstra, Chon, and Gu]{zijlstra2009five}
Zijlstra,~P.; Chon,~J.~W.; Gu,~M. Five-dimensional optical recording mediated by surface plasmons in gold nanorods. \emph{nature} \textbf{2009}, \emph{459}, 410--413\relax
\mciteBstWouldAddEndPuncttrue
\mciteSetBstMidEndSepPunct{\mcitedefaultmidpunct}
{\mcitedefaultendpunct}{\mcitedefaultseppunct}\relax
\EndOfBibitem
\bibitem[Khawas \latin{et~al.}(2024)Khawas, Bhattacharjee, Mukherjee, Sain, and Srivastava]{khawas2024directing}
Khawas,~S.; Bhattacharjee,~S.; Mukherjee,~S.; Sain,~A.; Srivastava,~S. Directing the formation of tunable superlattice crystalline phases from anisotropic nanoparticles. \emph{Colloids and Surfaces A: Physicochemical and Engineering Aspects} \textbf{2024}, \emph{690}, 133762\relax
\mciteBstWouldAddEndPuncttrue
\mciteSetBstMidEndSepPunct{\mcitedefaultmidpunct}
{\mcitedefaultendpunct}{\mcitedefaultseppunct}\relax
\EndOfBibitem
\bibitem[Kruse \latin{et~al.}(2024)Kruse, Rao, S{\'a}nchez-Iglesias, Monta{\~n}o-Priede, Iturrospe~Ibarra, Lopez, Seifert, Arbe, and Grzelczak]{kruse2024temperature}
Kruse,~J.; Rao,~A.; S{\'a}nchez-Iglesias,~A.; Monta{\~n}o-Priede,~J.~L.; Iturrospe~Ibarra,~A.; Lopez,~E.; Seifert,~A.; Arbe,~A.; Grzelczak,~M. Temperature-Modulated Reversible Clustering of Gold Nanorods Driven by Small Surface Ligands. \emph{Chemistry--A European Journal} \textbf{2024}, \emph{30}, e202302793\relax
\mciteBstWouldAddEndPuncttrue
\mciteSetBstMidEndSepPunct{\mcitedefaultmidpunct}
{\mcitedefaultendpunct}{\mcitedefaultseppunct}\relax
\EndOfBibitem
\bibitem[Zhang and Lin(2014)Zhang, and Lin]{zhang2014high}
Zhang,~Z.; Lin,~M. High-yield preparation of vertically aligned gold nanorod arrays via a controlled evaporation-induced self-assembly method. \emph{Journal of Materials Chemistry C} \textbf{2014}, \emph{2}, 4545--4551\relax
\mciteBstWouldAddEndPuncttrue
\mciteSetBstMidEndSepPunct{\mcitedefaultmidpunct}
{\mcitedefaultendpunct}{\mcitedefaultseppunct}\relax
\EndOfBibitem
\bibitem[Kim \latin{et~al.}(2014)Kim, Na, Ham, and Min]{kim2014mediating}
Kim,~Y.-K.; Na,~H.-K.; Ham,~S.; Min,~D.-H. Mediating ordered assembly of gold nanorods by controlling droplet evaporation modes for surface enhanced Raman scattering. \emph{RSC Advances} \textbf{2014}, \emph{4}, 50091--50096\relax
\mciteBstWouldAddEndPuncttrue
\mciteSetBstMidEndSepPunct{\mcitedefaultmidpunct}
{\mcitedefaultendpunct}{\mcitedefaultseppunct}\relax
\EndOfBibitem
\bibitem[Li \latin{et~al.}(2016)Li, Luo, Zhang, Li, Xiong, Qiao, Cao, Wang, He, and Jing]{li2016direct}
Li,~H.; Luo,~H.; Zhang,~Z.; Li,~Y.; Xiong,~B.; Qiao,~C.; Cao,~X.; Wang,~T.; He,~Y.; Jing,~G. Direct observation of nanoparticle multiple-ring pattern formation during droplet evaporation with dark-field microscopy. \emph{Physical Chemistry Chemical Physics} \textbf{2016}, \emph{18}, 13018--13025\relax
\mciteBstWouldAddEndPuncttrue
\mciteSetBstMidEndSepPunct{\mcitedefaultmidpunct}
{\mcitedefaultendpunct}{\mcitedefaultseppunct}\relax
\EndOfBibitem
\bibitem[Wang \latin{et~al.}(2010)Wang, Wang, He, Lv, and Wang]{wang2010preparation}
Wang,~R.-M.; Wang,~B.-Y.; He,~Y.-F.; Lv,~W.-H.; Wang,~J.-F. Preparation of composited Nano-TiO2 and its application on antimicrobial and self-cleaning coatings. \emph{Polymers for Advanced Technologies} \textbf{2010}, \emph{21}, 331--336\relax
\mciteBstWouldAddEndPuncttrue
\mciteSetBstMidEndSepPunct{\mcitedefaultmidpunct}
{\mcitedefaultendpunct}{\mcitedefaultseppunct}\relax
\EndOfBibitem
\bibitem[Nikoobakht and El-Sayed(2003)Nikoobakht, and El-Sayed]{nikoobakht2003preparation}
Nikoobakht,~B.; El-Sayed,~M.~A. Preparation and growth mechanism of gold nanorods (NRs) using seed-mediated growth method. \emph{Chemistry of materials} \textbf{2003}, \emph{15}, 1957--1962\relax
\mciteBstWouldAddEndPuncttrue
\mciteSetBstMidEndSepPunct{\mcitedefaultmidpunct}
{\mcitedefaultendpunct}{\mcitedefaultseppunct}\relax
\EndOfBibitem
\bibitem[Munief \latin{et~al.}(2018)Munief, Heib, Hempel, Lu, Schwartz, Pachauri, Hempelmann, Schmitt, and Ingebrandt]{munief2018silane}
Munief,~W.-M.; Heib,~F.; Hempel,~F.; Lu,~X.; Schwartz,~M.; Pachauri,~V.; Hempelmann,~R.; Schmitt,~M.; Ingebrandt,~S. Silane deposition via gas-phase evaporation and high-resolution surface characterization of the ultrathin siloxane coatings. \emph{Langmuir} \textbf{2018}, \emph{34}, 10217--10229\relax
\mciteBstWouldAddEndPuncttrue
\mciteSetBstMidEndSepPunct{\mcitedefaultmidpunct}
{\mcitedefaultendpunct}{\mcitedefaultseppunct}\relax
\EndOfBibitem
\bibitem[Berry \latin{et~al.}(2015)Berry, Neeson, Dagastine, Chan, and Tabor]{berry2015measurement}
Berry,~J.~D.; Neeson,~M.~J.; Dagastine,~R.~R.; Chan,~D.~Y.; Tabor,~R.~F. Measurement of surface and interfacial tension using pendant drop tensiometry. \emph{Journal of colloid and interface science} \textbf{2015}, \emph{454}, 226--237\relax
\mciteBstWouldAddEndPuncttrue
\mciteSetBstMidEndSepPunct{\mcitedefaultmidpunct}
{\mcitedefaultendpunct}{\mcitedefaultseppunct}\relax
\EndOfBibitem
\bibitem[Ahmad \latin{et~al.}(2019)Ahmad, Derkink, Boulogne, Bampoulis, Zandvliet, Khan, Jan, and Kooij]{ahmad2019self}
Ahmad,~I.; Derkink,~F.; Boulogne,~T.; Bampoulis,~P.; Zandvliet,~H.~J.; Khan,~H.~U.; Jan,~R.; Kooij,~E.~S. Self-assembly and wetting properties of gold nanorod--CTAB molecules on HOPG. \emph{Beilstein journal of nanotechnology} \textbf{2019}, \emph{10}, 696--705\relax
\mciteBstWouldAddEndPuncttrue
\mciteSetBstMidEndSepPunct{\mcitedefaultmidpunct}
{\mcitedefaultendpunct}{\mcitedefaultseppunct}\relax
\EndOfBibitem
\bibitem[Kwiecinski \latin{et~al.}(2019)Kwiecinski, Segers, Van Der~Werf, Van~Houselt, Lohse, Zandvliet, and Kooij]{kwiecinski2019evaporation}
Kwiecinski,~W.; Segers,~T.; Van Der~Werf,~S.; Van~Houselt,~A.; Lohse,~D.; Zandvliet,~H.~J.; Kooij,~S. Evaporation of dilute sodium dodecyl sulfate droplets on a hydrophobic substrate. \emph{Langmuir} \textbf{2019}, \emph{35}, 10453--10460\relax
\mciteBstWouldAddEndPuncttrue
\mciteSetBstMidEndSepPunct{\mcitedefaultmidpunct}
{\mcitedefaultendpunct}{\mcitedefaultseppunct}\relax
\EndOfBibitem
\bibitem[Shao \latin{et~al.}(2020)Shao, Duan, Hou, and Zhong]{shao2020role}
Shao,~X.; Duan,~F.; Hou,~Y.; Zhong,~X. Role of surfactant in controlling the deposition pattern of a particle-laden droplet: Fundamentals and strategies. \emph{Advances in colloid and interface science} \textbf{2020}, \emph{275}, 102049\relax
\mciteBstWouldAddEndPuncttrue
\mciteSetBstMidEndSepPunct{\mcitedefaultmidpunct}
{\mcitedefaultendpunct}{\mcitedefaultseppunct}\relax
\EndOfBibitem
\bibitem[Lee \latin{et~al.}(2008)Lee, Ivanova, Starov, Hilal, and Dutschk]{lee2008kinetics}
Lee,~K.; Ivanova,~N.; Starov,~V.; Hilal,~N.; Dutschk,~V. Kinetics of wetting and spreading by aqueous surfactant solutions. \emph{Advances in colloid and interface science} \textbf{2008}, \emph{144}, 54--65\relax
\mciteBstWouldAddEndPuncttrue
\mciteSetBstMidEndSepPunct{\mcitedefaultmidpunct}
{\mcitedefaultendpunct}{\mcitedefaultseppunct}\relax
\EndOfBibitem
\bibitem[Johansson \latin{et~al.}(2022)Johansson, Galli{\'e}ro, and Legendre]{johansson2022molecular}
Johansson,~P.; Galli{\'e}ro,~G.; Legendre,~D. How molecular effects affect solutal Marangoni flows. \emph{Physical Review Fluids} \textbf{2022}, \emph{7}, 064202\relax
\mciteBstWouldAddEndPuncttrue
\mciteSetBstMidEndSepPunct{\mcitedefaultmidpunct}
{\mcitedefaultendpunct}{\mcitedefaultseppunct}\relax
\EndOfBibitem
\bibitem[Schneider \latin{et~al.}(2012)Schneider, Rasband, and Eliceiri]{schneider2012nih}
Schneider,~C.~A.; Rasband,~W.~S.; Eliceiri,~K.~W. NIH Image to ImageJ: 25 years of image analysis. \emph{Nature methods} \textbf{2012}, \emph{9}, 671--675\relax
\mciteBstWouldAddEndPuncttrue
\mciteSetBstMidEndSepPunct{\mcitedefaultmidpunct}
{\mcitedefaultendpunct}{\mcitedefaultseppunct}\relax
\EndOfBibitem
\bibitem[McLintock \latin{et~al.}(2013)McLintock, Lee, and Wark]{mclintock2013stabilized}
McLintock,~A.; Lee,~H.~J.; Wark,~A.~W. Stabilized gold nanorod--dye conjugates with controlled resonance coupling create bright surface-enhanced resonance Raman nanotags. \emph{Physical Chemistry Chemical Physics} \textbf{2013}, \emph{15}, 18835--18843\relax
\mciteBstWouldAddEndPuncttrue
\mciteSetBstMidEndSepPunct{\mcitedefaultmidpunct}
{\mcitedefaultendpunct}{\mcitedefaultseppunct}\relax
\EndOfBibitem
\bibitem[Yue \latin{et~al.}(2020)Yue, Liu, Yan, and Jiang]{yue2020hierarchical}
Yue,~X.; Liu,~X.; Yan,~N.; Jiang,~W. Hierarchical Colloidosomes with a Highly Ordered and Oriented Arrangement of Gold Nanorods via Confined Assembly at the Emulsion Interface. \emph{The Journal of Physical Chemistry C} \textbf{2020}, \emph{124}, 20458--20468\relax
\mciteBstWouldAddEndPuncttrue
\mciteSetBstMidEndSepPunct{\mcitedefaultmidpunct}
{\mcitedefaultendpunct}{\mcitedefaultseppunct}\relax
\EndOfBibitem
\bibitem[Xie \latin{et~al.}(2013)Xie, Guo, Guo, He, Chen, Ji, Chen, Wu, Liu, and Xie]{xie2013controllable}
Xie,~Y.; Guo,~S.; Guo,~C.; He,~M.; Chen,~D.; Ji,~Y.; Chen,~Z.; Wu,~X.; Liu,~Q.; Xie,~S. Controllable two-stage droplet evaporation method and its nanoparticle self-assembly mechanism. \emph{Langmuir} \textbf{2013}, \emph{29}, 6232--6241\relax
\mciteBstWouldAddEndPuncttrue
\mciteSetBstMidEndSepPunct{\mcitedefaultmidpunct}
{\mcitedefaultendpunct}{\mcitedefaultseppunct}\relax
\EndOfBibitem
\bibitem[Staniscia \latin{et~al.}(2022)Staniscia, Guzman, and Kanduc]{staniscia2022tuning}
Staniscia,~F.; Guzman,~H.~V.; Kanduc,~M. Tuning contact angles of aqueous droplets on hydrophilic and hydrophobic surfaces by surfactants. \emph{The Journal of Physical Chemistry B} \textbf{2022}, \emph{126}, 3374--3384\relax
\mciteBstWouldAddEndPuncttrue
\mciteSetBstMidEndSepPunct{\mcitedefaultmidpunct}
{\mcitedefaultendpunct}{\mcitedefaultseppunct}\relax
\EndOfBibitem
\end{mcitethebibliography}


\providecommand{\latin}[1]{#1}
\makeatletter
\providecommand{\doi}
  {\begingroup\let\do\@makeother\dospecials
  \catcode`\{=1 \catcode`\}=2 \doi@aux}
\providecommand{\doi@aux}[1]{\endgroup\texttt{#1}}
\makeatother
\providecommand*\mcitethebibliography{\thebibliography}
\csname @ifundefined\endcsname{endmcitethebibliography}  {\let\endmcitethebibliography\endthebibliography}{}
\begin{mcitethebibliography}{9}
\providecommand*\natexlab[1]{#1}
\providecommand*\mciteSetBstSublistMode[1]{}
\providecommand*\mciteSetBstMaxWidthForm[2]{}
\providecommand*\mciteBstWouldAddEndPuncttrue
  {\def\EndOfBibitem{\unskip.}}
\providecommand*\mciteBstWouldAddEndPunctfalse
  {\let\EndOfBibitem\relax}
\providecommand*\mciteSetBstMidEndSepPunct[3]{}
\providecommand*\mciteSetBstSublistLabelBeginEnd[3]{}
\providecommand*\EndOfBibitem{}
\mciteSetBstSublistMode{f}
\mciteSetBstMaxWidthForm{subitem}{(\alph{mcitesubitemcount})}
\mciteSetBstSublistLabelBeginEnd
  {\mcitemaxwidthsubitemform\space}
  {\relax}
  {\relax}

\bibitem[Verhoeven(1996)]{verhoeven1996glossary}
Verhoeven,~J. Glossary of terms used in photochemistry (IUPAC Recommendations 1996). \emph{Pure and Applied Chemistry} \textbf{1996}, \emph{68}, 2223--2286\relax
\mciteBstWouldAddEndPuncttrue
\mciteSetBstMidEndSepPunct{\mcitedefaultmidpunct}
{\mcitedefaultendpunct}{\mcitedefaultseppunct}\relax
\EndOfBibitem
\bibitem[Schneider \latin{et~al.}(2012)Schneider, Rasband, and Eliceiri]{schneider2012nih}
Schneider,~C.~A.; Rasband,~W.~S.; Eliceiri,~K.~W. NIH Image to ImageJ: 25 years of image analysis. \emph{Nature methods} \textbf{2012}, \emph{9}, 671--675\relax
\mciteBstWouldAddEndPuncttrue
\mciteSetBstMidEndSepPunct{\mcitedefaultmidpunct}
{\mcitedefaultendpunct}{\mcitedefaultseppunct}\relax
\EndOfBibitem
\bibitem[Li and Neumann(1992)Li, and Neumann]{li1992contact}
Li,~D.; Neumann,~A. Contact angles on hydrophobic solid surfaces and their interpretation. \emph{Journal of colloid and interface science} \textbf{1992}, \emph{148}, 190--200\relax
\mciteBstWouldAddEndPuncttrue
\mciteSetBstMidEndSepPunct{\mcitedefaultmidpunct}
{\mcitedefaultendpunct}{\mcitedefaultseppunct}\relax
\EndOfBibitem
\bibitem[Wu and Brzozowski(1971)Wu, and Brzozowski]{wu1971surface}
Wu,~S.; Brzozowski,~K.~J. Surface free energy and polarity of organic pigments. \emph{Journal of Colloid and Interface Science} \textbf{1971}, \emph{37}, 686--690\relax
\mciteBstWouldAddEndPuncttrue
\mciteSetBstMidEndSepPunct{\mcitedefaultmidpunct}
{\mcitedefaultendpunct}{\mcitedefaultseppunct}\relax
\EndOfBibitem
\bibitem[Owens and Wendt(1969)Owens, and Wendt]{owens1969estimation}
Owens,~D.~K.; Wendt,~R. Estimation of the surface free energy of polymers. \emph{Journal of applied polymer science} \textbf{1969}, \emph{13}, 1741--1747\relax
\mciteBstWouldAddEndPuncttrue
\mciteSetBstMidEndSepPunct{\mcitedefaultmidpunct}
{\mcitedefaultendpunct}{\mcitedefaultseppunct}\relax
\EndOfBibitem
\bibitem[Popov(2005)]{popov2005evaporative}
Popov,~Y.~O. Evaporative deposition patterns: spatial dimensions of the deposit. \emph{Physical Review E—Statistical, Nonlinear, and Soft Matter Physics} \textbf{2005}, \emph{71}, 036313\relax
\mciteBstWouldAddEndPuncttrue
\mciteSetBstMidEndSepPunct{\mcitedefaultmidpunct}
{\mcitedefaultendpunct}{\mcitedefaultseppunct}\relax
\EndOfBibitem
\bibitem[McLintock \latin{et~al.}(2013)McLintock, Lee, and Wark]{mclintock2013stabilized}
McLintock,~A.; Lee,~H.~J.; Wark,~A.~W. Stabilized gold nanorod--dye conjugates with controlled resonance coupling create bright surface-enhanced resonance Raman nanotags. \emph{Physical Chemistry Chemical Physics} \textbf{2013}, \emph{15}, 18835--18843\relax
\mciteBstWouldAddEndPuncttrue
\mciteSetBstMidEndSepPunct{\mcitedefaultmidpunct}
{\mcitedefaultendpunct}{\mcitedefaultseppunct}\relax
\EndOfBibitem
\bibitem[Yue \latin{et~al.}(2020)Yue, Liu, Yan, and Jiang]{yue2020hierarchical}
Yue,~X.; Liu,~X.; Yan,~N.; Jiang,~W. Hierarchical Colloidosomes with a Highly Ordered and Oriented Arrangement of Gold Nanorods via Confined Assembly at the Emulsion Interface. \emph{The Journal of Physical Chemistry C} \textbf{2020}, \emph{124}, 20458--20468\relax
\mciteBstWouldAddEndPuncttrue
\mciteSetBstMidEndSepPunct{\mcitedefaultmidpunct}
{\mcitedefaultendpunct}{\mcitedefaultseppunct}\relax
\EndOfBibitem
\end{mcitethebibliography}

\end{document}

\newpage

\section{Characterization of AuNRs} \label{sec:AuNR-char}

The synthesized gold nanorods are characterized using UV-visible spectroscopy, Transmission Electron Microscopy (TEM) imaging, and Zeta potential measurement. 

\subsection{UV-Visible spectroscopy}
\begin{figure}[h!]
\centering
    \includegraphics[width={0.5\textwidth}]{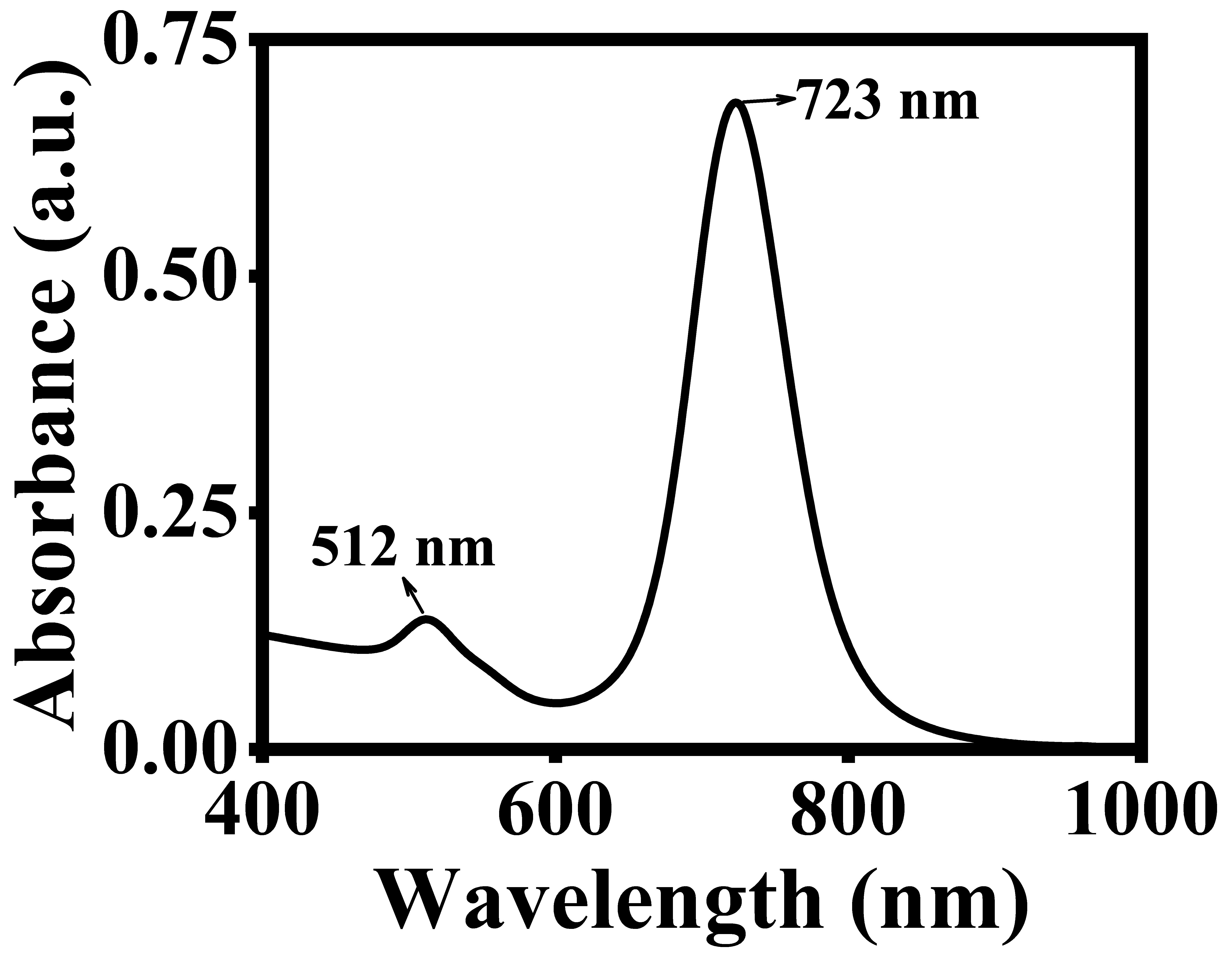}
    \caption{UV-Visible spectra for AuNRs showing LSPR and TSPR peaks}
    \label{fig-UV}
\end{figure}

UV-visible spectra for AuNR and AuNP colloidal solution are obtained using JASCO V-730
Spectrophotometer. For AuNR [Figure \ref{fig-UV}], we observe two peaks for longitudinal (LSPR
at 723 nm) and transverse (TSPR at 512 nm) plasmon resonance due to the anisotropic geometry. Calculation for concentration of AuNR solution is done using Beer-Lambert’s law\cite{verhoeven1996glossary} from the LSPR absorbance. 

\subsection{TEM micrograph and size analysis}

\begin{figure}[h!]
\centering
    \includegraphics[width={0.9\textwidth}]{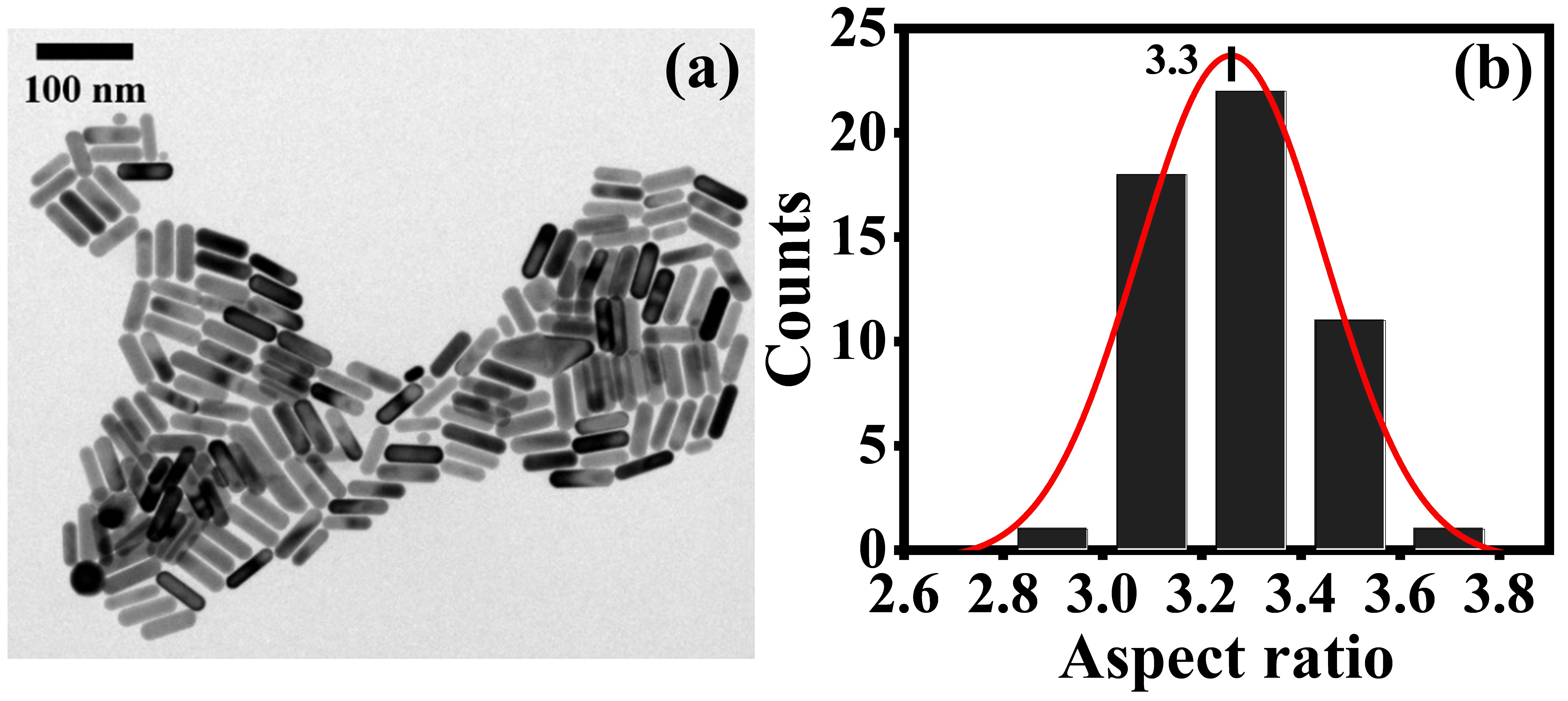}
    \caption{TEM micrograph (a) and estimation of the aspect ratio (b) of the synthesized AuNRs are shown here. An average aspect ratio of 3.3 is obtained from the Gaussian distribution.}
    \label{fig-TEM}
\end{figure}

Transmission electron microscopy (TEM) micrographs showing monodisperse AuNRs in Fig. \ref{fig-TEM}a. Estimation of length (L) and width (D) of the rods were done using ImageJ\cite{schneider2012nih} software to calculate the aspect ratio (L/D). The aspect ratio distribution shows an average of $\sim~3.3$ for the AuNRs. 

\subsection{Zeta potential measurement}
\begin{figure}[h!]
\centering
    \includegraphics[width={0.5\textwidth}]{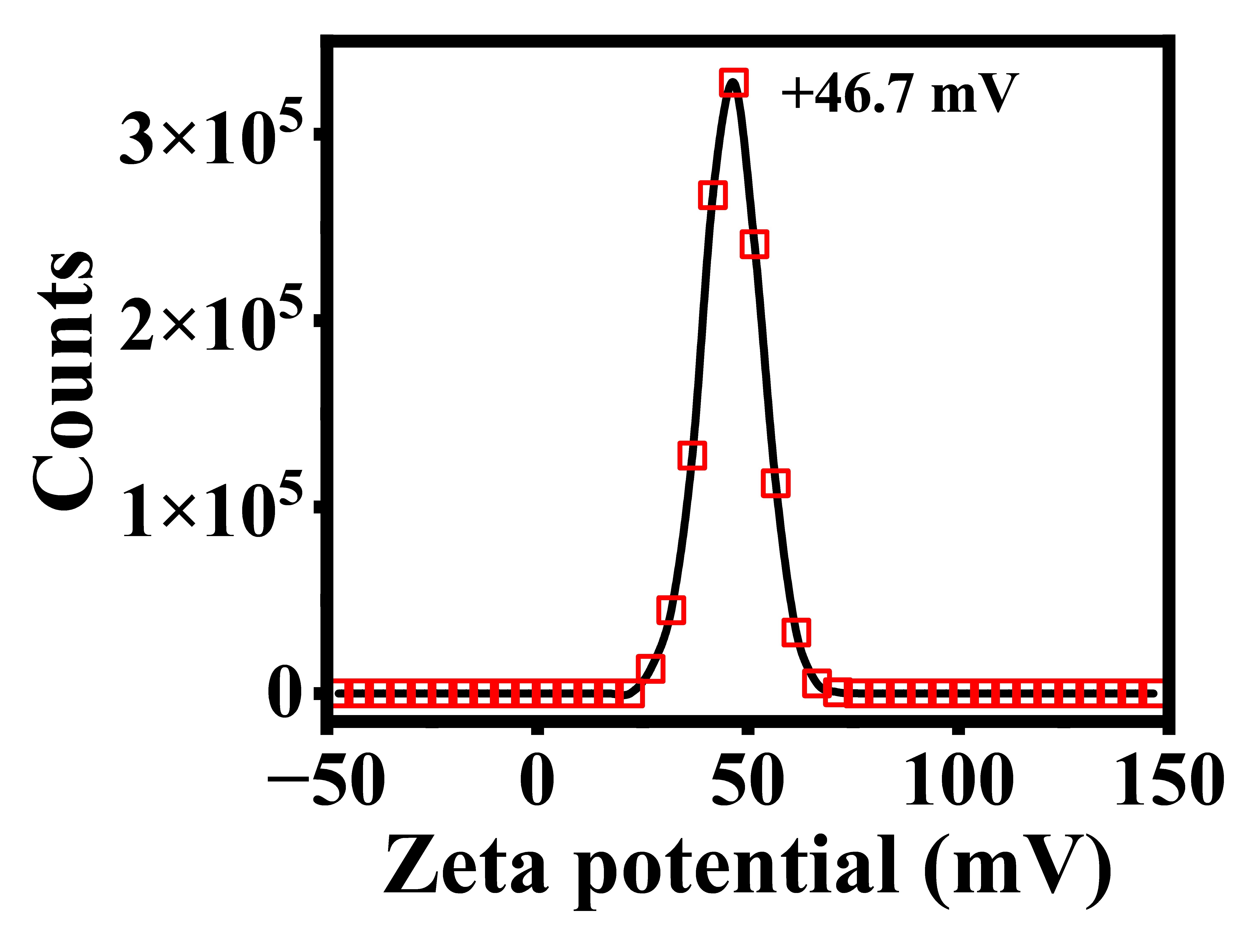}
    \caption{Zeta potential measurement of AuNRs showing a net positive surface charge.}
    \label{fig-Zeta}
\end{figure}

The Zeta potential of the AuNR solution was measured using the Zetasizer NanoZS instrument (Malvern). Fig.\ref{fig-Zeta} shows a high positive ($\sim+47~mV$) value of zeta potential for the nanorods, which suggests a net positive surface coverage due to CTAB coating.

\section{Characterization of hydrophobic surface} \label{sec:Si-char}

The PFOTS-coated Si substrate is characterized using surface free energy (SFE) measurement and atomic force microscopy (AFM) measurement to account for its hydrophobic nature ($\gamma_{SV}$ and roughness).

\subsection{SFE measurement}

SFE measurement or the estimation of $\gamma_{SV}$ for the hydrophobic surface has been done using the optical tensiometer. Sessile drops of a polar (water) and non-polar (Di-iodomethane) liquid are used to find the contact angle ($\theta$). 

\begin{table}[h!]
    \centering
    \begin{tabular}{|c|c|c|c|c|c|}
    \hline
     Systems/Parameters & $\gamma_{LV}^{tot}~(mN/m)$  &  $\gamma^+~(mN/m)$ & $\gamma^-~(mN/m)$ & $\gamma^d~(mN/m)$ & $\theta~(^o)$\\
     \hline
    Water & 72.8 & 25.5& 25.5 & 21.8 &108.38\\
    \hline
    Di-iodomethane & 50.8 & 0& 0& 50.8&92.21\\
    \hline            
    \end{tabular}
    \caption{Values for different parameters for the calculation of SFE of PFOTS-coated Si substrate}
    \label{tab:SFE}
\end{table}

$\gamma_{SV}$ is calculated using the equation of state by Neumann et al.\cite{li1992contact}, the harmonic mean method by Wu et al.\cite{wu1971surface}, and the OWRK method\cite{owens1969estimation}. The average $\gamma_{SV}$ is found to be $16.5\pm 2.5~mN/m$, which signifies the enhanced hydrophobicity of the surface.

\subsection{AFM measurement}

\begin{figure}[h!]
\centering
    \includegraphics[width={0.9\textwidth}]{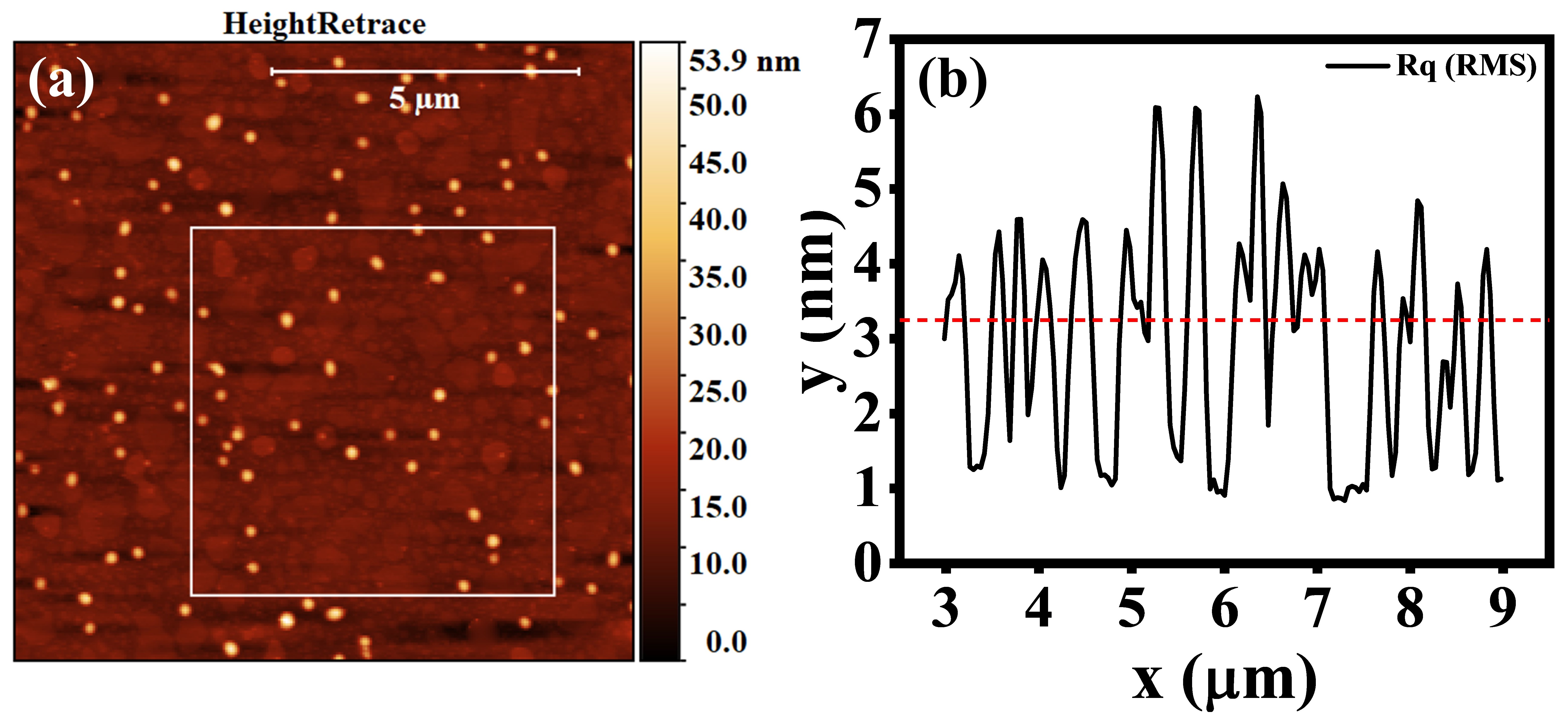}
    \caption{(a) Surface morphology of PFOTS-coated Si surface from AFM measurement and (b) RMS roughness profile corresponding to the boxed region.}
    \label{fig-kpFM}
\end{figure}

To estimate the surface roughness due to hydrophobic coating on the Si substrate, Atomic Force Microscopy (AFM) is done. $10\times10~\mu m^2$ scans of the substrate showing the height sensor profile (a) and the roughness map (b) are given here. From Fig.\ref{fig-kpFM}a (boxed region), the RMS roughness ($R_q$) profile is plotted in Fig.\ref{fig-kpFM}b. Average surface roughness is found to be $\sim 3.25\pm1.37~nm$.

\section{Micro-scale variations in deposit patterns}
Low-magnification SEM micrographs of dried depositions are shown for $~2~nM-30~nM$. The coffee-ring to disc-shaped transition is visible beyond $7.5~nM$. Also, the ratio of onset of deposition ($t_d$) to $T_{max}$ is also shown for each image, confirming that nanorod deposition occurs at late stages of drying, for all the concentrations.

\begin{figure}[h!]
\centering
    \includegraphics[width={0.9\textwidth}]{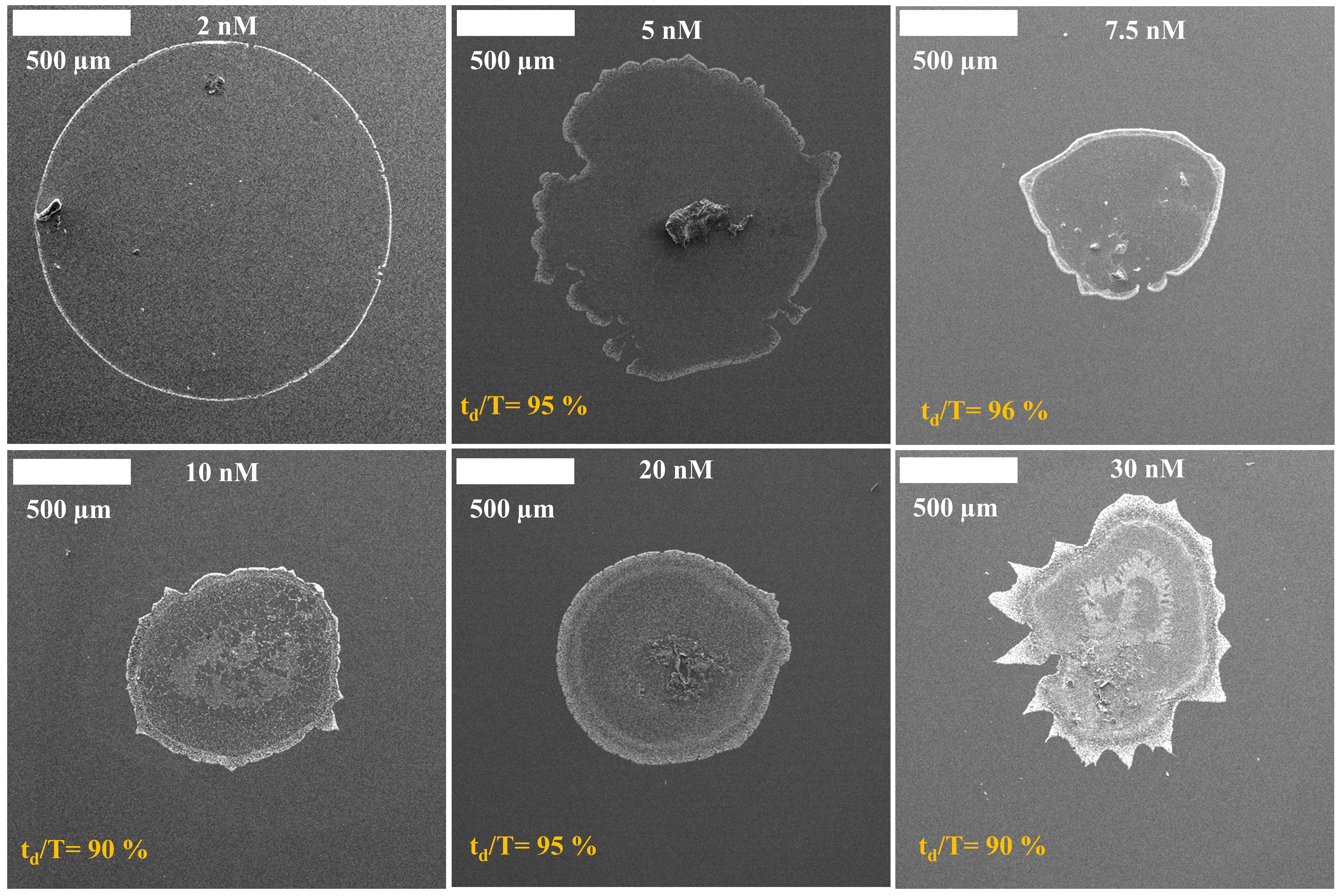}
    \caption{Dried patterns formed by AuNR droplets on the hydrophobic substrate with variation in concentration are shown here. The transition from coffee-ring to disc-shaped deposition is observed as we increase AuNR concentration.}
    \label{fig:all_lowmag}
\end{figure}

\section{Variation in $D_0$ and $D_{SEM}$ with concentration}
The difference in initial droplet diameter ($D_{0}$) and the statistical diameter of the dried deposits ($D_{SEM}$) shown in \ref{fig:all_lowmag} calculated using ImageJ software is tabulated below. The difference in $D_{0}$ and $D_{SEM}$ also justifies that particle deposition occurs at the later stage of evaporation, even if the contact line is strongly receded.

\begin{table}[h!]
    \centering
    \begin{tabular}{|c|c|c|}
    \hline
     Concentration (nM) & $D_0$ (mm)  &  $D_{SEM}$ (mm)\\
    \hline
        2 & 1.53 & 1.52\\
    \hline    
        5 & 1.62 & 1.38\\
    \hline    
        7.5 & 1.62 & 0.85
    \\\hline    
        10 & 1.73 & 0.97\\
    \hline    
        20 & 2.1 & 1.01\\
    \hline    
        30 & 2.0 & 1.26\\
    \hline
    \end{tabular}
    \caption{Variation of $D_{0}$ and $D_{SEM}$ with concentration }
    \label{tab:D0-DSEM}
\end{table}

\section{Coffee ring width (CRW) variation with AuNR concentration}
\label{sec:popov}
To understand the scaling of normalized coffee-ring width (CRW) with concentration\cite{popov2005evaporative}, the ratio of statistical width of coffee-ring (calculated using ImageJ software) to $D_{SEM}$ is plotted. 
\begin{figure}[h!]
\centering
    \includegraphics[width={0.5\textwidth}]{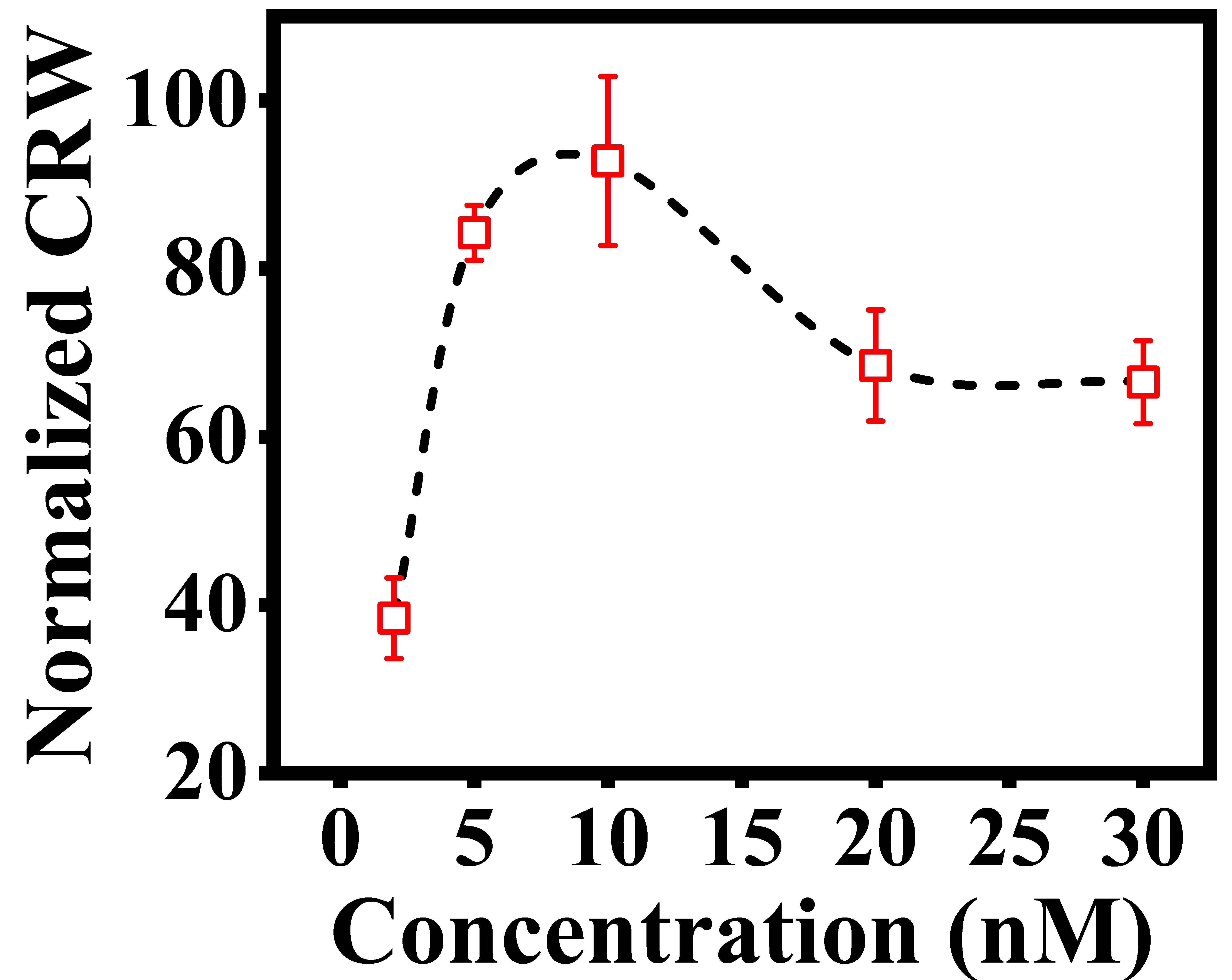}
    \caption{Variation in normalized CRW with AuNR concentration}
    \label{fig:CRW-vs-C}
\end{figure}
According to Popov's model\cite{popov2005evaporative}, normalized CRW is expected to scale as $C^{0.5}$. But, in our system, CRW decreases beyond $7.5~nM$ and then saturates.

\section{Preparation of AuNR-Nile red conjugates for Confocal microscopy}

A hydrophobic dye can be sequestered into the CTAB bilayer, forming nanorod-dye conjugate\cite{mclintock2013stabilized}. The stock solution of Nile red was made by dissolving a few milligrams of dye in Dimethyl sulfoxide (DMSO). The dye was added to AuNR filled aliquots such that the initial nanorod concentrations were $2~nM$ and $30~nM$, and the dye concentration was $5~\mu M$. The mixture was incubated for $16~hrs$ under ambient conditions and centrifuged thrice at $4774g$ for $20~mins$ to remove the excess amount of dye present in the solvent. Nile red is a hydrophobic dye and gets attached to the hydrophobic part of the CTAB bilayer\cite{yue2020hierarchical}. A $560~nm$ DPSS Laser is used for Confocal microscopy experiments.

\section{Area analysis of vertically oriented AuNR }
\begin{figure}[h!]
\centering
    \includegraphics[width={0.5\textwidth}]{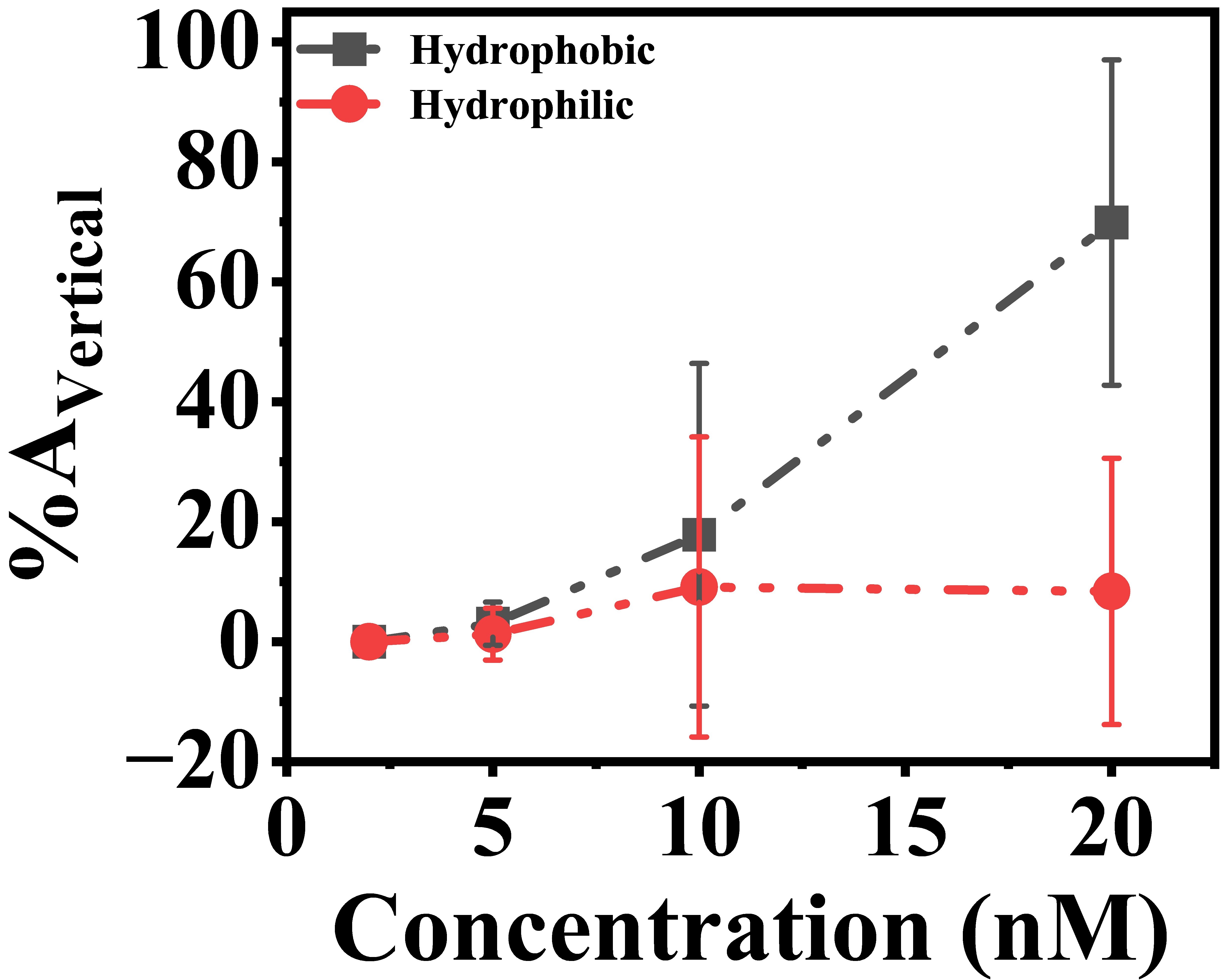}
    \caption{Percentage area of vertically oriented AuNRs on hydrophobic (black) and hydrophilic (red) substrates for different concentrations.}
    \label{Fig-vertical area}
\end{figure}
The percentage area occupied by vertically oriented AuNRs from SEM micrographs is calculated using ImageJ software for $2~nM$ to $20~nM$. SEM micrographs from coffee-ring to the inner parts were taken radially for all the samples and the vertically occupied area was calculated. In the case of the hydrophilic substrate, AuNRs are mostly ordered smectically. \par

\begin{figure}[h!]
\centering
    \includegraphics[width={0.9\textwidth}]{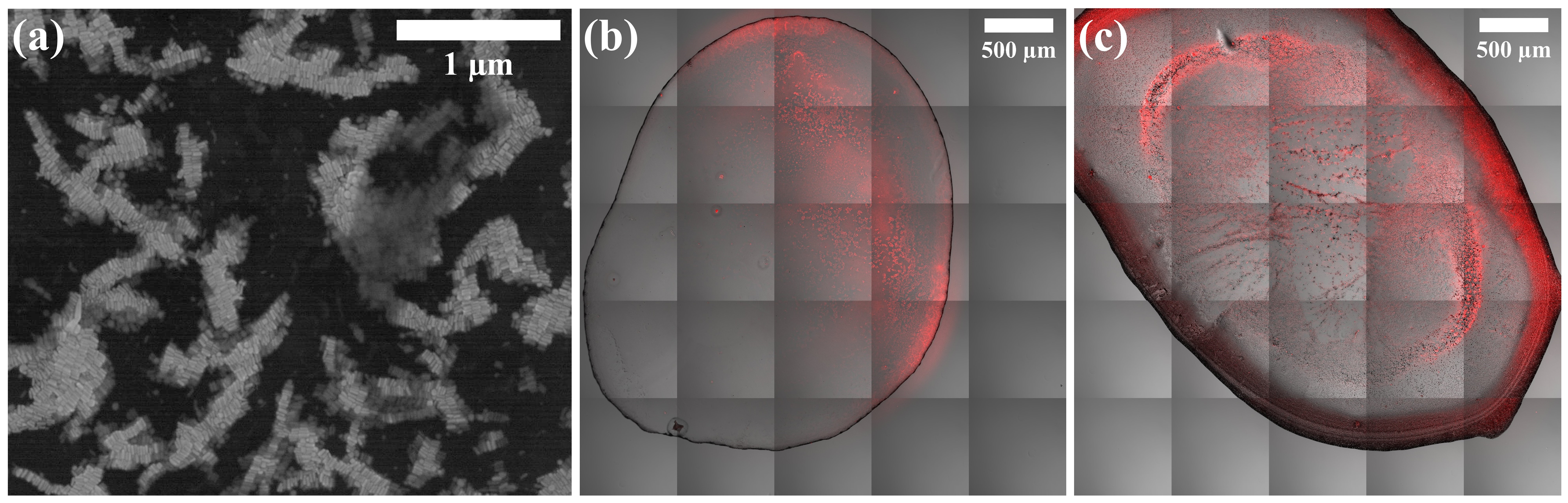}
    \caption{(a) SEM micrographs of $20~nM$ AuNR on the hydrophilic surface near the inner vicinity of the coffee-ring. The confocal microscopy images of (b) $2~nM$ and (c) $20~nM$ Nile red-tagged AuNR on hydrophilic coverslips.}
    \label{Fig-hydrophilic data}
\end{figure}
High-magnification SEM micrographs of $20~nM$ AuNR on hydrophilic $Si$ - substrate ($1:1$ piranha cleaned) are shown in Fig. \ref{Fig-hydrophilic data}a. Here, AuNRs forms isotropic/smectic clusters near the inner vicinity of the coffee-ring. The deposition of CTAB on the surface is probed by using Nile-red-tagged AuNR and the confocal microscopy images [\ref{Fig-hydrophilic data}(b,c)] show CTAB is mostly deposited near the coffee-ring for both low and high-concentrations of AuNR. 
\newpage
\bibliography{references}